\documentclass[twocolumn]{aastex631}
\usepackage{amsmath}
\usepackage{enumitem}
\usepackage{booktabs}
\usepackage{makecell}
\usepackage{xcolor}
\newcommand{\mytilde}{\raise.19ex\hbox{$\scriptstyle\sim$}}
\newcommand{\qcr}[1]{{\fontfamily{qcr}\selectfont #1}}
\newcommand{\bs}{\boldsymbol}

\shorttitle{Weak-Lensing Analysis of the Double Radio-relic Cluster PSZ2\,G181.06+48.47}
\shortauthors{Ahn et al.}

\begin{document}

\title{PSZ2\,G181.06+48.47 III: weak-lensing analysis and merging scenario reconstruction \\ of a low-mass cluster with exceptionally-distant radio relics}

\correspondingauthor{M. James Jee}
\email{eunmo.ahn@yonsei.ac.kr, mkjee@yonsei.ac.kr}

\author[0009-0009-4676-7868]{Eunmo Ahn}
\affiliation{Department of Astronomy, Yonsei University, 50 Yonsei-ro, Seoul 03722, Republic of Korea}

\author[0000-0001-5966-5072]{Hyejeon Cho}
\affiliation{Department of Astronomy, Yonsei University, 50 Yonsei-ro, Seoul 03722, Republic of Korea}
\affiliation{Center for Galaxy Evolution Research, Yonsei University, 50 Yonsei-ro, Seoul 03722, Republic of Korea}

\author[0000-0002-5751-3697]{M. James Jee}
\affiliation{Department of Astronomy, Yonsei University, 50 Yonsei-ro, Seoul 03722, Republic of Korea}
\affiliation{Department of Physics and Astronomy, University of California, Davis, One Shields Avenue, Davis, CA 95616, USA}

\author[0000-0002-1566-5094]{Wonki Lee}
\affiliation{Department of Astronomy, Yonsei University, 50 Yonsei-ro, Seoul 03722, Republic of Korea}

\author[0000-0001-8322-4162]{Andra Stroe}
\affiliation{Center for Astrophysics \textbar\ Harvard \& Smithsonian, 60 Garden St., Cambridge, MA 02138, USA}
\affiliation{Space Telescope Science Institute, 3700 San Martin Drive, Baltimore, MD 21218, USA}

\author[0000-0001-7509-2972]{Kamlesh Rajpurohit}
\affiliation{Center for Astrophysics \textbar\ Harvard \& Smithsonian, 60 Garden St., Cambridge, MA 02138, USA}

\author[0000-0002-4462-0709]{Kyle Finner}
\affiliation{IPAC, California Institute of Technology, 1200 E California Blvd., Pasadena, CA 91125, USA}

\author[0000-0002-9478-1682]{William Forman}
\affiliation{Center for Astrophysics \textbar\ Harvard \& Smithsonian, 60 Garden St., Cambridge, MA 02138, USA}

\author[0000-0003-2206-4243]{Christine Jones}
\affiliation{Center for Astrophysics \textbar\ Harvard \& Smithsonian, 60 Garden St., Cambridge, MA 02138, USA}

\author[0000-0002-0587-1660]{Reinout van Weeren}
\affiliation{Leiden Observatory, Leiden University, PO Box 9513, 2300 RA Leiden, The Netherlands}

\begin{abstract}
The galaxy cluster PSZ2\,G181.06+48.47 ($z=0.234$) is a post-merging system that exhibits symmetric double radio relics separated by \mytilde2.7~Mpc.
We present the first weak-lensing analysis of PSZ2\,G181.06+48.47 and propose possible merging scenarios using numerical simulations.
Our analysis with Subaru Hyper Suprime-Cam imaging identifies a binary dark matter structure consisting of northern and southern components, separated by \mytilde500~kpc.
Assuming Navarro-Frenk-White (NFW) halos, the masses for the northern and southern subclusters are $M_{200c}^\text{N} = 0.88_{-0.30}^{+0.35} \times 10^{14} M_{\odot}$ and $M_{200c}^\text{S} = 2.71_{-0.48}^{+0.51} \times 10^{14} M_{\odot}$, respectively.
By superposing the two NFW halos, we determine the total mass of the cluster to be $M_{200c} = 4.22_{-1.00}^{+1.10} \times 10^{14} M_{\odot}$ ($M_{500c} = 2.90_{-0.69}^{+0.75} \times 10^{14} M_{\odot}$).
Our mass estimate suggests that the two relics are located around the cluster $R_{200c}$, where the density of the intracluster medium is very low.
Our idealized simulations find that an off-axis collision of a 3:1 major merger can simultaneously reproduce the observed relic and dark matter halo separations.
From these findings, we suggest that the system is observed \mytilde0.9~Gyr after the first pericenter passage and is returning from the first apocenter.
\end{abstract}

\keywords{Galaxy clusters(584) - Weak gravitational lensing(1797) - Radio continuum emission(1340)}

\section{Introduction}\label{sec:introduction}
Galaxy clusters are the largest gravitationally bound systems in the Universe.
They grow in mass primarily through major mergers, which release gravitational energy on the order of $10^{64}$ erg \citep[e.g.,][]{Markevitch.1999, Ricker.2001}, the most energetic events since the Big Bang.
Cluster mergers span several gigayears, making it essential to study their merger histories to understand the evolution of galaxy clusters and the surrounding large-scale structures.

During cluster mergers, shocks are generated when the collision velocity exceeds the sound speed of the intracluster medium (ICM).
Merger shocks dissipate \mytilde30\% of the gravitational energy into the ICM \citep[e.g.,][]{Sarazin.2002}.
While the exact mechanisms are still under debate and may vary depending on specific properties of the individual clusters, it is known that merger shocks amplify the ICM magnetic fields and accelerate cosmic-ray electrons (CRe) to relativistic speeds, with Lorentz factors exceeding $\gamma > 10^3$ \citep[e.g.,][]{Gabici.2003, Iapichino.2012, Pinzke.2013, Hong.2014, Porter.2015}.

\begin{figure*}
    \centering
    \includegraphics[width=0.9\textwidth]{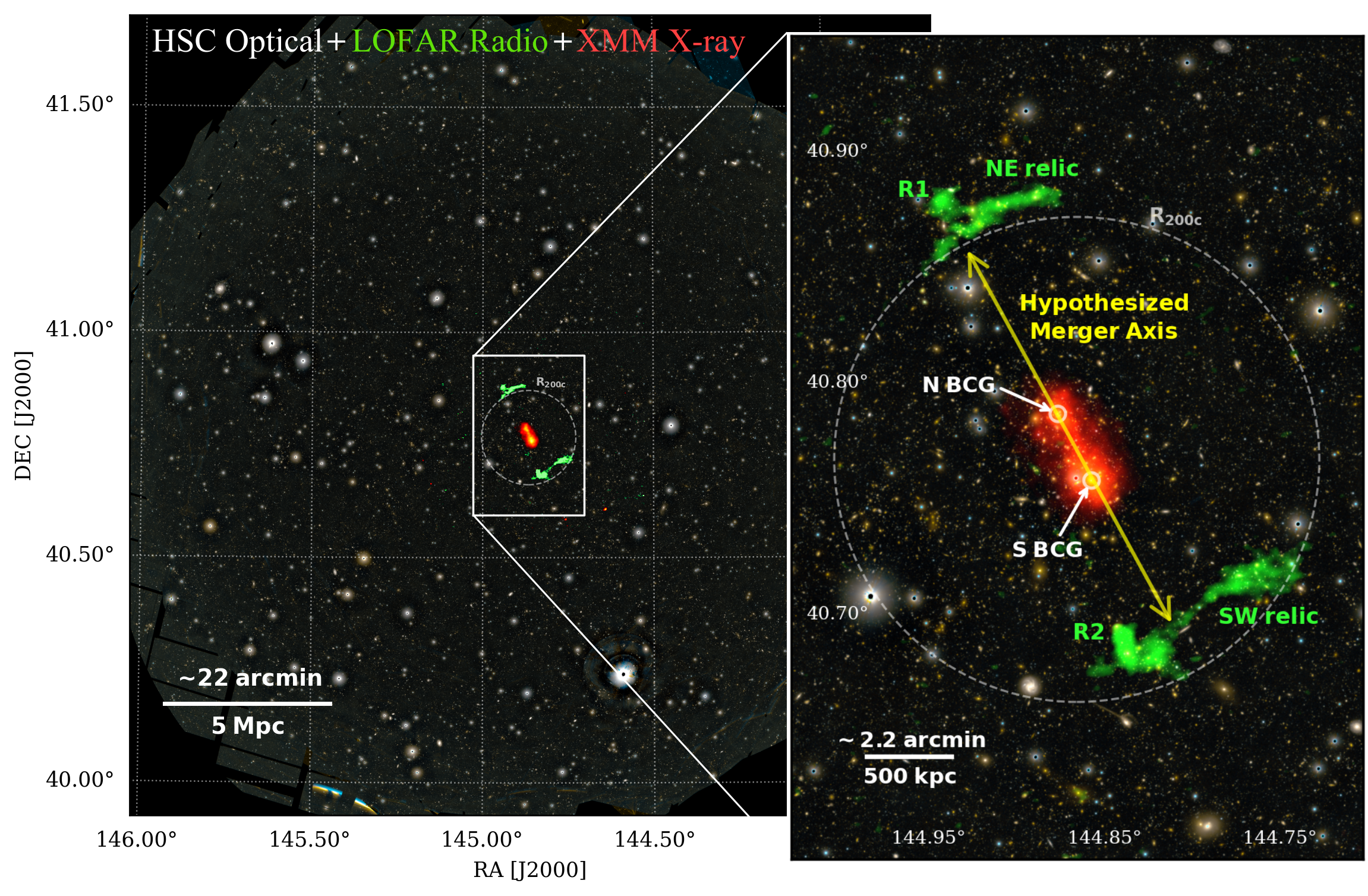}
    \caption{Multi-wavelength view of the PSZ2G181 field. 
    The background false-color image was created by combining the Subaru Hyper Suprime-Cam $g$- and $r2$-band images.
    Red and green colors represent the XMM-Newton X-ray \citep[][]{Stroe.2025} and LOFAR 150~MHz radio \citep[][]{Shimwell.2022} emission intensities, respectively. The left figure shows the $\mytilde2.5\ \text{deg}^2$ entire field of view of our mosaic image, and the right zoom-in figure shows the central 3~Mpc $\times$ 5~Mpc region. The hypothesized merger axis is defined as a vector connecting two BCGs, which is parallel to a vector connecting the two relics.}
    \label{fig:color_xray_radio}
\end{figure*}
In the presence of magnetic fields, the accelerated CRe emit diffuse synchrotron radio emission in the outskirts of colliding clusters, known as `radio relics' \citep[see][for a review and references therein]{Feretti.2012, Brunetti.2014, vanWeeren.2019}.
Radio relics serve as tracers of merger shocks \citep[e.g.,][]{Ensslin.1998}, which are not being decelerated by the cluster gravitational potential \citep[e.g.,][]{Ha.2018}.
Therefore, radio relics offer crucial insights into the merger history of clusters, including parameters such as time since collision, viewing angle, collision velocity, etc. \citep[e.g.,][]{Ng.2015, Golovich.2019a, Wonki.2020, Finner.2024}.

In principle, every cluster merger generates one or more shockwaves traveling outward from the cluster core.
However, in practice, only a small fraction of merging clusters exhibit radio relics.
For merger shocks to be detected as radio relics, several conditions must be met.
First, the collision should occur near the plane of the sky; otherwise, the radio emission would be fainter than the edge-on-viewed relic and may fall below the detection limit of observations \citep[e.g.,][]{Skillman.2013}.
Second, the collision velocity must be high enough to produce powerful shocks with high Mach numbers.
As a result, cosmic ray electrons are accelerated to relativistic energies, enabling them to emit synchrotron radiation at radio frequencies.
Additionally, radio relics are only observable during a limited time window \citep[e.g.,][]{Nuza.2024}.
Radio relics are hard to form right after the collision as the kinetic power through the shock surface is low near the cluster center \citep[][]{Vazza.2012}.
On the other hand, the CRe density is too low in the outermost regions of the cluster, making it difficult to produce bright relics there.
Consequently, radio relics are rare among merging clusters.
For example, among the \mytilde300 clusters selected from the LOw-Frequency ARray (LOFAR) Two-meter Sky Survey \citep[LoTSS-DR2;][]{Shimwell.2022, Jones.2023}, only about \mytilde10\% host one or more radio relic candidates.

Symmetric double radio relics, positioned opposite each other equidistantly from the cluster center, are observed even less frequently among radio relic systems.
Since the two relics trace the same merger event, double radio relic systems are ideal astrophysical laboratories for probing the cluster merger including dynamical histories, particle acceleration, large-scale structure formation, dark matter (DM) properties, etc. \citep[e.g.,][]{Rottgering.1997, vanWeeren.2011, Stroe.2013, Stuardi.2019, Finner.2021, Cho.2022, Koribalski.2024}.
Among the aforementioned LoTSS-DR2 radio relic clusters, only six exhibit symmetric double radio relics positioned opposite each other relative to the cluster center.

The mass of colliding DM halos is crucial for analyzing cluster mergers.
DM halos account for \mytilde85\% of the total mass of clusters, predominantly influencing the merger trajectories.
Moreover, observations provide only a single snapshot of the merging process that spans billions of years. 
Therefore, accurately mapping and identifying DM subhalos and estimating their masses are essential for setting the initial conditions in numerical simulations of individual mergers.
These steps enable the robust reconstruction of their merging scenarios.

Weak gravitational lensing (WL) is a powerful tool for measuring cluster masses and mapping the DM distribution within galaxy clusters \citep[e.g.,][]{Ragozzine.2012, Jee.2014_520, Monteiro-Oliveira.2020, Ahn.2024, Finner.2024}.
The method is not entirely bias-free but is unaffected by the cluster's dynamical state since it solely uses the geometry of the observer, lens, and background sources.
This advantage becomes particularly valuable for merging clusters, which are dynamically unstable.
For instance, if cluster member galaxies are not in virial equilibrium, mass estimates based on velocity dispersion can be biased \citep[e.g.,][]{Pinkney.1996, Takizawa.2010}.
Additionally, because of the ongoing merger activities, the intracluster medium (ICM) might not be in a relaxed state, which can similarly affect the X-ray or Sunyaev–Zel'dovich (SZ) mass estimates \citep[e.g.,][]{Rasia.2006, Krause.2012}.

In this paper, we present the first WL analysis of the merging cluster PSZ2\,G181.06+48.47 (PSZ2G181 hereafter) at $z=0.234$ (with equatorial coordinates $\mathrm{09^h 39^m 28^s}$, $\mathrm{40^d 45^m 59^s}$) and propose possible merging scenarios utilizing numerical simulations.
This cluster is remarkable for its symmetric double radio relics, which have an unusually large projected separation of \mytilde2.7~Mpc, as well as the symmetrical distribution of its brightest cluster galaxies (BCGs) and X-ray emission.
Figure~\ref{fig:color_xray_radio} provides the multi-wavelength view of the cluster, showing the two distinct BCGs and the elongated X-ray emission extending between them.
The two prominent radio relics, positioned equidistant from the midpoint of the two BCGs, indicate that this cluster is in a post-merger state.
This is likely the result of a binary collision between the two subclusters.

PSZ2G181 was identified as a galaxy cluster using the redMaPPer algorithm \citep[RM~J093930.4+404710.4;][]{Rykoff.2014} and the SZ effect with an estimated mass of $M_{\mathrm{500c,SZ}} = 4.2 \pm 0.5 \times 10^{14} M_{\odot}$ \citep[][]{Planck.2016}.
While this initial SZ detection did not reveal any notable features, \cite{Botteon.2022} recently identified two diffuse radio sources located at the outskirts of the cluster from LoTSS-DR2.
Both sources have been confirmed as radio relics through deep, high-quality radio data across a broad frequency range of 120~MHz to 1.5~GHz \citep{Rajpurohit.2025}.
The \mytilde2.7~Mpc separation is the largest known to date among the double radio relic systems with the mass range of $M_{500} < 5 \times 10^{14} M_{\odot}$ \citep[][]{Stroe.2025}.
The lower mass of colliders results in a slower collision speed, which in turn leads to slower shock speeds \citep[e.g.,][]{Springel.2007}.
As a result, the large relic separation observed in the low mass cluster PSZ2G181 suggests that this system is in an extremely late stage of the merger.

The properties of the two relics do not significantly deviate from known relations such as the relic power-cluster mass relation \citep[][]{Jones.2023, Duchesne.2024, Stroe.2025}.
However, the unusual combination of the large separation of the relics and the relatively low mass of the cluster makes this system particularly interesting.
It raises important questions about the particle acceleration mechanisms in low-density environments, especially regarding how the relics can maintain their bright luminosity even near the virial radius of the cluster.

To fully dissect this unique system, we initiated the multi-wavelength project of this cluster, including optical, radio, and X-ray analyses.
As a part of this project, we analyzed the Subaru/Hyper Suprime-Cam data and identified the DM substructures, quantified the DM halo masses, and suggested the possible merging scenario of this cluster.
\cite{Rajpurohit.2025} proposed the physics and formation mechanisms of the non-thermal diffuse radio sources utilizing the deep uGMRT and JVLA data, while \cite{Stroe.2025} analyzed the $\mathit{Chandra}$ and XMM-Newton data and revealed the X-ray discontinuities including the thermodynamical properties of the cluster.

This paper is organized as follows.
The Subaru observations and data reduction are described in Section \ref{sec:observations}.
The overall analysis, including the basic WL theory, PSF modeling, shape measurement, source selection, and redshift estimation are presented in Section \ref{sec:WL}.
We provide the results of mass reconstruction and mass estimation in Section \ref{sec:results}, and we propose the possible merging scenarios using numerical simulations in Section \ref{sec:discussion}.

We assume a flat $\mathrm{\Lambda CDM}$ cosmology with $H_0\mathrm{ = 70~km~s^{-1}~Mpc ^{-1}}$ and $\mathrm{\Omega_M}=0.3$.
At the cluster redshift ($z=0.234$), the angular size of $1''$ corresponds to the physical size of \mytilde3.7~kpc.
Cluster masses $M_{\mathrm{200c}}$ ($M_{\mathrm{500c}}$) are defined as the mass enclosed by a sphere of radius $R_{\mathrm{200c}}$ ($R_{\mathrm{500c}}$), where the average density is 200 (500) times the critical density at the cluster redshift.
All magnitudes are in the AB magnitude system.
All errors are quoted at the 1-$\sigma$ level unless otherwise noted.

\section{Observations}\label{sec:observations}
\subsection{Hyper Suprime-Cam Imaging}\label{sub:HSC}
We observed the PSZ2G181 field with the Subaru Hyper Suprime-Cam
\citep[HSC;][]{Furusawa.2018, Kawanomoto.2018, Komiyama.2018, Miyazaki.2018} on 2023 January 15, 16, and 22 (PI: H. Cho).
HSC is the wide-field imaging camera composed of 104 science CCDs with a pixel scale of $0\farcs168\,\text{pixel}^{-1}$, covering a $\mytilde1.5^{\circ}$ diameter of the field of view ($\mytilde20$~Mpc at the cluster redshift).
We observed the cluster with the HSC-$g$ and HSC-$r2$ bands that bracket the 4000 $\mathrm{\AA}$ break at the cluster redshift.
We applied field rotation and dithering across all 12 exposures in both the HSC-$g$ and HSC-$r2$ bands, significantly reducing artifacts such as cosmic rays, diffraction spikes, saturation trails, etc.
For the $r2$-band observations, we discarded exposures that were taken under relatively poor seeing conditions ($\text{FWHM}\gtrsim 1\farcs0$).
Including these poor-seeing images decreases the overall background rms noise, however, it also smooths the shape of the background galaxies that contain the lensing signals.
The final mean seeings are $1\farcs07$ and $0\farcs67$ for the $g$ and $r2$ bands, respectively.
The total integrations of the final mosaic images are 2040 and 3240~s for the $g$ and $r2$ bands, respectively.

\subsection{Data Reduction}\label{sub:reduction}
Single-frame calibration including overscan/bias/dark subtraction, flat fielding, astrometric calibration, etc, were processed with the LSST Science Pipelines\footnote{https://pipelines.lsst.io/index.html} stack v26\_0\_0 \citep{Bosch.2018, Bosch.2019, Jenness.2022}.
This pipeline returns the astrometric solutions in the Simple Imaging Polynomial (SIP) convention.
Since we used the {\tt SWarp} software \citep[][]{Bertin.2002} to create the final mosaic images, we converted the SIP convention to the {\tt SWarp}-readable TPV projection type \citep[e.g,][]{Finner.2023, HyeongHan.2024a} using the {\tt sip\_tpv} code\footnote{https://github.com/stargaser/sip\_tpv} \citep[][]{Shupe.2012}.
The final deep mosaic images where the WL signal is measured were created by inverse-variance weight-averaging using {\tt SWarp}, excluding outliers that deviate more than $3\sigma$ from the median-stacked images.

We ran {\tt SExtractor} \citep[][]{Bertin.1996} in dual-image mode, using the $r2$-band image for detection because it is deeper and has a sharper point-spread function (PSF) than the $g$-band image.
Since the $g$-band PSF is \mytilde1.6 times larger than that of the $r2$-band, the summation of the flux within the isophotal area defined from the $r2$-band image underestimates the $g$-band flux.
Therefore, we used \qcr{MAG\_AUTO} both for the total magnitude and color estimation for both filters.
We calibrated the photometry using the SDSS DR16 catalog \citep[][]{Ahumada.2020} covering the same field.
The $5\sigma$ limiting magnitudes for the point sources are 26.9 and 27.5~mag for $g$ and $r2$, respectively.

\section{Weak Lensing Analysis}\label{sec:WL}
\subsection{Theoretical Background}
WL measures weak distortion of background galaxies caused by foreground lenses.
Although the intrinsic shapes of background galaxies are unknown, their average distortions reveal the projected shear field caused by a foreground lens (e.g., a galaxy cluster).
In this section, we provide a brief overview of the theoretical background.
Readers are referred to review papers \citep[e.g.,][]{Mellier.1999, Bartelmann.2001, Schneider.2006} for details.

The true angular positions of the background galaxies $\bs{\beta}$ are mapped into the observed locations $\bs{\theta}$ by lensing.
The coordinate mapping from the source plane $\bs{\beta}$ to the image plane $\bs{\theta}$ can be linearized with the Jacobian matrix $\bs{A}$:
\begin{equation}\label{eq:jacobian}
    \bs{A} = \frac{\partial\bs{\beta}}{\partial\bs{\theta}} = (1-\kappa) \begin{pmatrix} 1-g_1 & -g_2 \\ -g_2 & 1+g_1 \end{pmatrix}.
\end{equation}
The convergence $\kappa$ represents the isotropic focusing of light rays (i.e., changes in the size of objects), while the reduced shear $g$ stands for the anisotropic focusing (i.e., distortions in shapes).
The reduced shear is defined as $g = \gamma / (1 - \kappa)$, where $\gamma$ is the shear.

$\kappa$ is a dimensionless surface mass density $\Sigma$ in units of the critical surface mass density $\Sigma_{cr}$:
\begin{equation}\label{eq:kappa}
    \kappa = \frac{\Sigma}{\Sigma_{cr}}, \;\;
    \Sigma_{cr} = \frac{c^2}{4 \pi G} \frac{D_s}{D_{l}D_{ls}},
\end{equation}
where $D_s$, $D_l$, and $D_{ls}$ are the angular diameter distances to the source, to the lens, and from the lens to the source, respectively.
The first component of the reduced shear $g_1$ distorts the image along the $x$- or $y$-axes, and the second component $g_2$ distorts the image along the $y=x$ and $y=-x$ directions.

The matrix $\bs{A}$ transforms a circle into an ellipse with the ellipticity $g=\sqrt{g_1^2+g_2^2}=(a-b)/(a+b)$, where $a$ and $b$ are the semimajor and semiminor axes, respectively.
The position angle $\phi$ of the ellipse is $\text{(1/2)tan}^{-1}(g_2/g_1)$.
The complex notation of the reduced shear can then be written as
\begin{equation}
    \bs{g}=g_1+\bs{i}g_2=ge^{2i\phi}.
\end{equation}

The local shear caused by the foreground lens transforms the intrinsic complex ellipticity $\bs{\epsilon}$ into the observed complex ellipticity $\bs{e}$ as
\begin{equation}\label{eq:ellip trans}
    \bs{e} = \frac{\bs{\epsilon} + \bs{g}}{1 + \bs{g}^* \bs{\epsilon}},
\end{equation}
where the asterisk (*) is the complex conjugate.
Assuming that galaxies are randomly oriented, i.e. $\left<\bs{\epsilon} \right>=0$, we can average the observed ellipticities to estimate the reduced shear as
\begin{equation}\label{eq:g_e}
    \bs{g} = \left< \bs{e} \right>.
\end{equation}
Therefore, we can use $\left<\bs{e}\right>$ as an estimator of the reduced shear when no systematic bias exists.

\subsection{PSF Modeling}
\begin{figure*}
    \includegraphics[width=\textwidth]{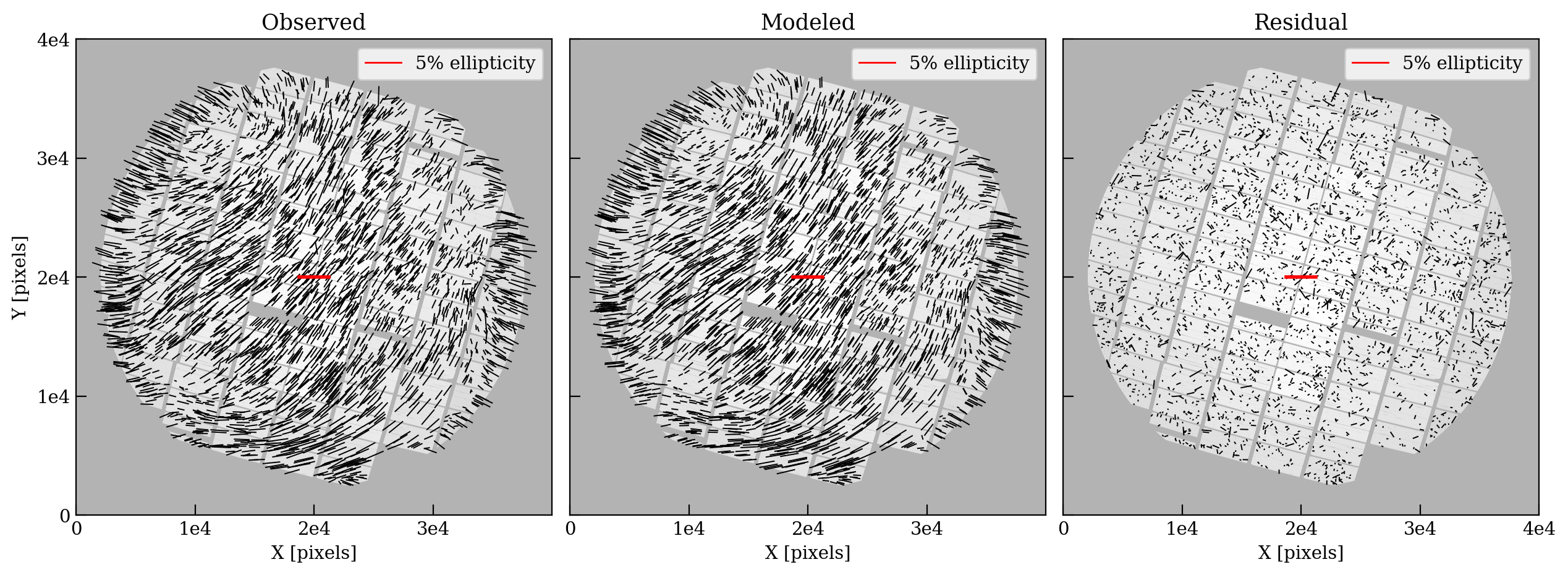}
    \caption{Example of the ``whisker'' plot measured from a single-visit HSC image overlaid onto its weight frame. The length and orientation of each whisker represent the size and direction of the PSF ellipticities, respectively. Our PSF model (middle) nicely reproduces the observed star patterns (left) across the entire focal plane including the edge of the field where the discontinuous changes of the PSF directions are observed. The residuals (right) between the observed and modeled stars are small.}
    
    \label{fig:r 8th whisker}
\end{figure*}
The observed shapes of galaxies are the result of the convolution of their true shapes with the PSF.
The size of the PSF dilutes the shear signal, while its elongation introduces the anisotropic bias. 
As a result, both the size and direction of the PSF 
impact the reconstruction of mass distribution and mass estimates.
Therefore, precise and accurate modeling of the PSF is one of the most critical steps in WL analysis.
We performed a principal component analysis \citep[PCA; see][for details]{Jee.2007, Jee.2011} for our PSF modeling.

As described in Section \ref{sub:reduction}, the mosaic images were created by outlier-clipped mean stacking of all contributing CCDs.
In the mean-stacked image, the PSF is equivalent to a linear combination of all frames used in the mosaic image.
Therefore, the PSF model should be created by precisely redoing the mosaic image co-adding procedure, using the same weights applied in the mosaic image.
Figure~\ref{fig:r 8th whisker} shows an example of the spatial variation of the $r2$-band PSF in our observation.
Although the PSF pattern flows smoothly except for the peripheral region, it is hard to reproduce the precise spatial variations across the CCDs using a single 2D polynomial function for the entire focal plane.
Furthermore, the PSF patterns on the stacked mosaic image have discontinuities across the CCD gaps \citep[e.g.,][]{Jee.2013}.
Therefore, we selected stars and performed PCA analysis for each CCD and stacked to obtain the PSF pattern in the mosaic image \citep[e.g.,][]{Jee.2015, Finner.2017, Yoon.2020}.

\begin{figure*}[ht]
    \centering
    \includegraphics[width=0.9\textwidth]{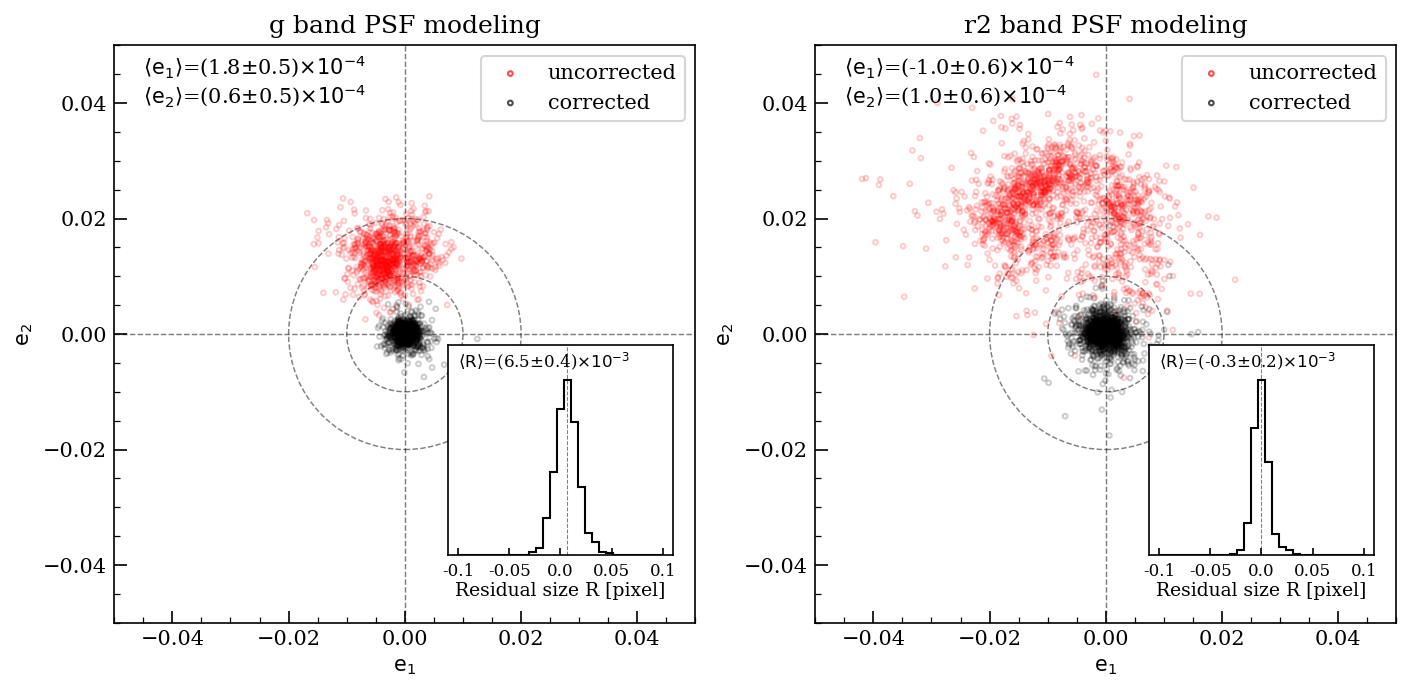}
    \caption{PSF correction results for the $g$-band (left) and $r2$-band (right) stars. The red and black circles represent the two ellipticity components of the stars before and after PSF correction, respectively. 
    The inset histograms show the residuals in size after subtracting the size of the PSF models from that of the stars. The average residuals of both the ellipticity components and size confirm the high fidelity of our PSF modeling.}
    \label{fig:psf residual}
\end{figure*}
We identified 20–40 high S/N stars (S/N $>$ 30) in each CCD for PSF modeling.
These stars are isolated from other objects, not saturated, and do not have zero-weight pixels.
The stars are then cropped into 21 pixels $\times$ 21 pixels postage stamps, applying subpixel shifting to position their peaks at the center of the central pixel.
Then, the coefficients of the principal components were fitted using 2D polynomial functions.
We fitted $3^{rd}$ ($2^{nd}$) order polynomials for the CCDs that have more than 30 (20) useful stars.
CCD frames that do not have a sufficient number of usable stars for PSF modeling were excluded from our image coadding\footnote{Among the total of 927 (1236) science frames in the $r2$ ($g$) band data, 11 (65) frames were excluded, mostly at the outermost part of the focal plane.}.
As mentioned before, the PSF model in the mosaic image was created via weight-averaging PSFs from all contributing CCDs.

To check whether our PSF models reproduce the observed star patterns well, we define the two ellipticity components $e_1$ and $e_2$ and the size of the stars $R$ as
\begin{equation}\label{eq:complex e star}
    e_1+\bs{i}e_2=\frac{Q_{11}-Q_{22}+2\bs{i}Q_{12}}{Q_{11}+Q_{22}+2(Q_{11}Q_{22}-Q_{12}^2)^{1/2}},
\end{equation}
\begin{equation}
    R=\sqrt{Q_{11}+Q_{22}}.
\end{equation}
$Q_{ij}$ ($i,j\in\{1,2 \}$) are the tensor of second brightness moments, which is defined as
\begin{equation}
    Q_{ij} = \frac{\int W(\bs{\theta}) I(\bs{\theta}) (\theta_i - \bar{\theta}_i) (\theta_j - \bar{\theta}_j) d^2 \bs{\theta}}
    {\int W(\bs{\theta}) I(\bs{\theta}) d^2 \bs{\theta}},
\end{equation}
where the weight function $W(\bs{\theta})$ is applied to suppress the noise in the outskirts, $I(\bs{\theta})$ is the pixel intensity at $\bs{\theta}$, and $\bar{\theta}_{i,j}$ is the center of the star.
Figure~\ref{fig:psf residual} shows the PSF modeling results for the $g$ and $r2$ bands.
The uncorrected $r2$-band stars have larger ellipticities than $g$-band stars because, in general, PSF becomes rounder as seeing increases. This happens because atmospheric turbulence is mostly isotropic, which masks the telescope's optical aberrations.
The residual ellipticity components ($e_1$, $e_2$) are centered on the origin with the reduced scatter for both bands.

Since we interpolated the observed star pattern to determine the PSF at each galaxy position, the polynomial fitting process may occasionally result in slight overfitting or underfitting.
\begin{figure}
    \includegraphics[width=\columnwidth]{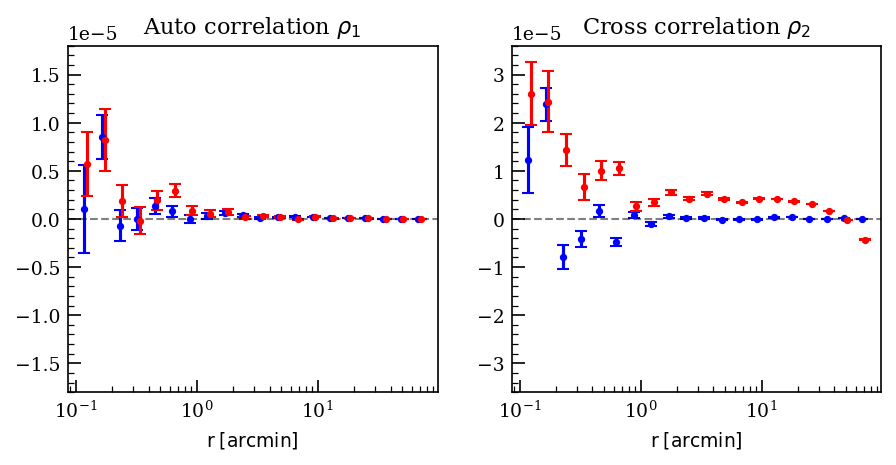}
    \caption{Ellipticity residual-residual autocorrelation (left) and data-residual cross-correlation (right) of the PSF modeling. Blue and red points represent the case for the $g$ and $r2$ bands, respectively. Amplitudes of both diagnostic functions are close to zero at the large radii, suggesting minimal residual systematics in our PSF model.}
    \label{fig:rowe}
\end{figure}
To investigate the presence of systematics in PSF modeling, we used the $\rho$ statistics proposed by \cite{Rowe.2010}:
\begin{align}
    \rho_1(r) &= \left<(\bs{e}-\bs{e_m})^*(\bs{e}-\bs{e_m}) \right> (r), \\
    \rho_2(r) &= \left<\bs{e}^*(\bs{e}-\bs{e_m}) + (\bs{e} - \bs{e_m})^*\bs{e} \right> (r)
\end{align}
where $\bs{e}$ and $\bs{e_m}$ are the complex ellipticities (Equation~\ref{eq:complex e star}) of the observed and modeled star, respectively.
The $\rho_1$ and $\rho_2$ statistics are diagnostic functions that measure the residual-residual autocorrelation and data-residual cross-correlation, respectively.
Figure~\ref{fig:rowe} shows the $\rho_1$ and $\rho_2$ diagnostics of the PSF stars in the PSZ2G181 field.
The amplitudes of both diagnostic functions are less than $10^{-5}$ at large angular separations, indicating that the systematics from the polynomial fitting process are negligible.
The small amplitude of $\rho_2$ at $r\mytilde10'$ ($\mytilde 5\times10^{-6}$) may result from minor additive bias.
However, we found that this bias contributes less than the order of magnitude of the shot noise.

\subsection{Shape Measurement}\label{sub:shape measurement}
We employed a forward-modeling approach for our shear measurement \citep[e.g.,][]{Jee.2016, Finner.2017, Cho.2022}.
All galaxies were cropped into square postage stamps for the $\chi^2$ minimization.
These stamps should contain enough pixels to mitigate truncation bias \citep[][]{Mandelbaum.2015}.
Other detections within the stamps were masked out to ensure that they did not contribute to the $\chi^2$ minimization.
We constructed the $\chi^2$ function to minimize the difference between the observed galaxy stamp $O$ and the PSF-convolved elliptical Gaussian function $M$ as:
\begin{equation}
    \chi^2 = \sum_{i}^{n} \sum_{j}^{g,r} \left(\frac{O_{i,j} - M_{i,j}}{\sigma_{i,j}} \right)^2,
\end{equation}
where $\sigma$ is the background rms noise.
The summations over $i$ and $j$ were taken for all $n$ pixels within the postage stamp and for the different filters, respectively.
Since the overall S/N is lower for the $g$ band galaxies than that in the $r2$ bands, detections with $\text{S/N}<5$ in the $g$ band images were not used in the fitting.
Employing multiple filters simultaneously for the minimization reduces the measurement errors of the ellipticity components and thereby increases the net S/N \citep[][]{Ahn.2024}.

The elliptical Gaussian function has seven free parameters: background and peak intensities ($I_{bkg}$ and $I_{peak}$), two centroids ($x_c$, $y_c$), $e_1$ and $e_2$, and semiminor axis $b$.
We define the $e_1$ and $e_2$ as:
\begin{align}
    e_1 &= e \cos(2\phi), \\
    e_2 &= e \sin(2\phi).
\end{align}
We fixed $I_{bkg}$ and ($x_c$, $y_c$) using \qcr{BACKGROUND} and (\qcr{XWIN\_IMAGE}, \qcr{YWIN\_IMAGE}) output by {\tt SExtractor}, respectively.
We used the {\tt MPFIT} code \citep[][]{Markwardt.2009} for the $\chi^2$ minimization, obtaining the solutions $e_1$, $e_2$ and $b$ for each galaxy.
$I_{peak}$ is different for each filter.

Analytic models are not perfect representations of real galaxies, leading to systematic errors \citep[model bias; e.g.,][]{Voigt.2010, Melchior.2010}.
Additionally, the non-linear relation between pixel and ellipticity introduces further systematic errors \citep[noise bias; e.g.,][]{Melchior.2012, Refregier.2012}.
Other systematics, such as selection bias and imperfect deblending, etc. also bias the true signal from background sources \citep[e.g.,][]{Jarvis.2016, Dawson.2016, FenechConti.2017, Mandelbaum.2018}.
Instead of addressing each of these systematic errors individually, we run image simulations that match observational data and derive shear calibration factors.
We obtained a global multiplicative factor of $m=1.15$ with a negligible additive bias \citep{Jee.2016}.
The multiplicative factor is applied to measured ellipticity components to derive the reduced shear as $g = m e$.
Readers are referred to \cite{Jee.2011} and \cite{Jee.2013} for details.

\subsection{Source Selection}
Foreground objects and cluster member galaxies are not lensed by the cluster and thereby dilute the lensing signal if they are included in the source catalog.
Additionally, the shape-fitting process becomes unstable for faint and small sources.
Therefore, careful selection of the source population is another critical step for WL analysis.
In this section, we describe our source selection criteria in terms of photometry and shape.

\subsubsection{Photometric Selection Criteria}\label{subsub:pho selection}
The optimal selection of background sources is to use galaxies with redshifts larger than the cluster redshift.
However, neither photometric nor spectroscopic redshift data are available for our field for the background source selection.
Therefore, we selected background sources using a color-magnitude relation, aiming to minimize contamination from foreground and cluster member galaxies and thereby maximize the net S/N.

In general, lensing S/N is maximized by carefully balancing the purity and the shot noise.
If we aggressively attempt to remove foreground galaxies to maximize purity, the number of background galaxies also decreases, which in turn increases the shot noise.
To estimate the net S/N in terms of the purity and the shot noise, we define the lensing S/N parameter as \citep[e.g.,][]{Jee.2017, Jinhyub.2019, HyeongHan.2024b}
\begin{equation}
    \mathrm{S/N_{lens}} = \eta \beta \sqrt{n},
\end{equation}
where $\eta$ is the purity of the source population, $\beta$ is the lensing efficiency, and $n$ is the number of sources.
We define the purity as
\begin{equation}
    \eta = 1 - (\text{foreground fraction}),
\end{equation}
and the lensing efficiency is defined as
\begin{equation}
    \beta = \int_{z_l}^{\infty}p(z)D_{ls} / D_s\;dz,
\end{equation}
where $z_l$ is the lens redshift and $p(z)$ is the probability distribution of the source redshift.
\begin{figure}[t]
    \includegraphics[width=\columnwidth]{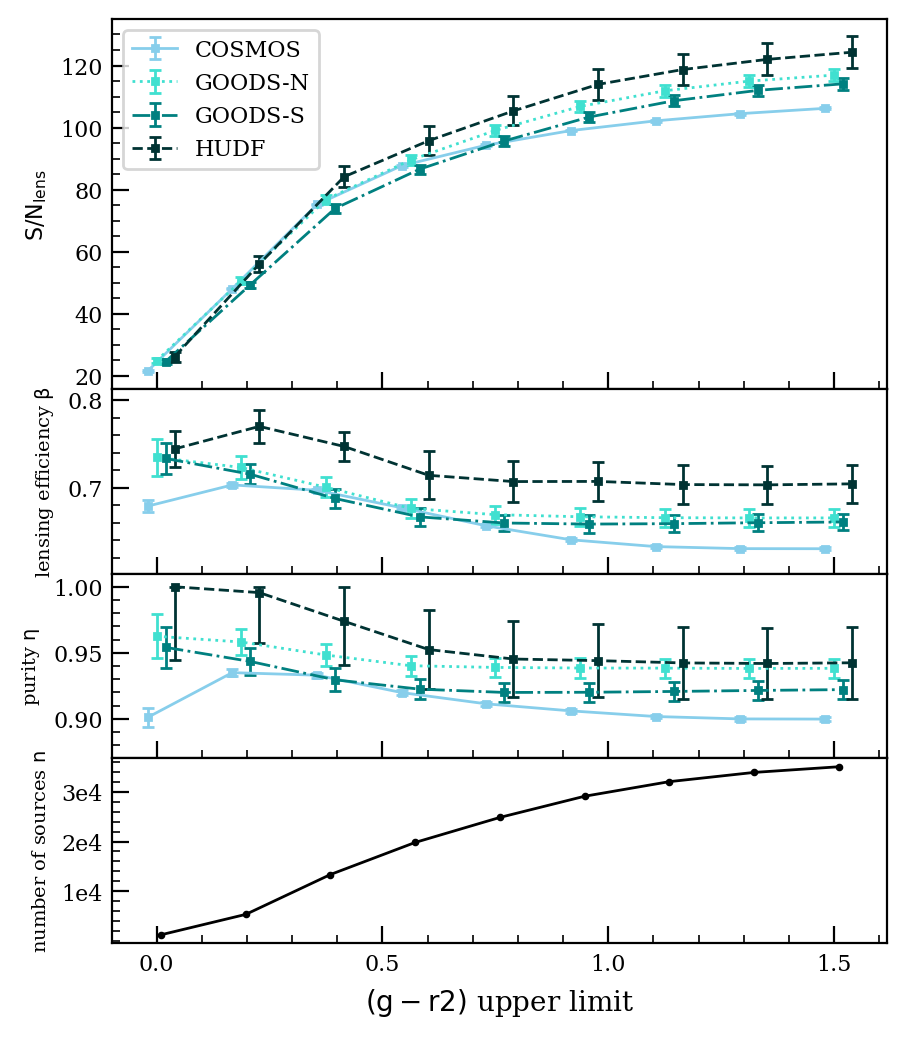}
    \caption{From top to bottom: $\mathrm{S/N_{lens}}$, lensing efficiency $\beta$, purity $\eta$, and number of sources $n$ as a function of the upper limit of the $(g-r2)$ color of the source population.
    The increase in number of sources for the higher $(g-r2)$ upper limit surpasses the decrement of $\beta$ and $\eta$, resulting in the increment of $\mathrm{S/N_{lens}}$.
    }
    \label{fig:lensing_SN}
\end{figure}
We estimated $\mathrm{S/N_{lens}}$ utilizing the control fields: Cosmic Evolution Survey \citep[COSMOS;][]{Weaver.2022}, Great Observatories Origins Deep Survey \citep[GOODS;][]{Kodra.2023}, and Hubble Ultra Deep Field \citep[HUDF;][]{Rafelski.2015}.
ACS magnitudes were transformed to HSC magnitudes when necessary.
Figure~\ref{fig:lensing_SN} shows the $\mathrm{S/N_{lens}}$ value and its parameters as a function of the upper limit of the $(g-r2)$ color of the source population.
The magnitude range is fixed to $23<r2<27$.
While the purity and lensing efficiency gradually decrease as the upper limit of the $(g - r2)$ color increases, the net $\mathrm{S/N_{lens}}$ value increases because of the reduction in shot noise.

However, as the cluster redshift decreases, the source population becomes less diluted by foreground galaxies, making cluster member contamination a more significant factor in its dilution (as shown in the third panel of Figure~\ref{fig:lensing_SN}, more than 90\% of the galaxies are expected to be located farther than the cluster redshift regardless of the reference field).
Also, previous studies have argued that the faint cluster member population can significantly dilute the lensing signal \citep[e.g.,][]{Broadhurst.2005, Okabe.2010, Melchior.2017, Varga.2019, Hamana.2020}.
This contamination cannot be estimated using control fields because the redshift distributions of galaxies in control fields differ from those in galaxy cluster fields.
One useful way to check for faint member contamination is to examine the cluster-centric radial distribution of the sources \citep[e.g.,][]{Hoekstra.2007, Applegate.2014, Jee.2017}.
\begin{figure}
    \includegraphics[width=\columnwidth]{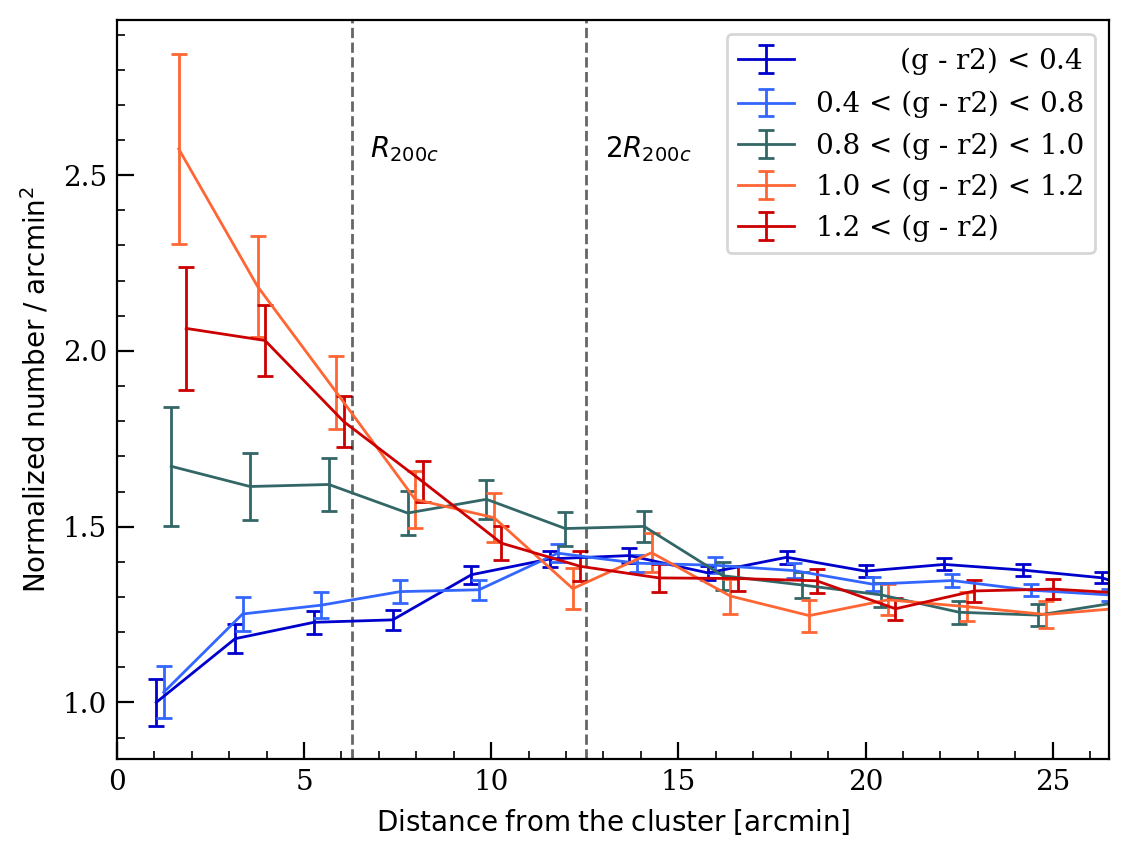}
    \caption{Radial profiles of the normalized source densities. The magnitude range is fixed to $23<r2<27$. The number density of the sources with $(g-r2) > 0.8$ shows an excess up to $r\sim2R_{200c}$. The profiles of the five different populations converge and become flat at this point.}
    \label{fig:radial_source}
\end{figure}
Figure~\ref{fig:radial_source} shows the radial distribution of the galaxies with varying color selection ranges.
Galaxies redder than $(g - r2) = 0.8$ clearly exhibit a central excess even at radii larger than the virial radius.
This suggests the existence of the faint red member galaxies, which also significantly dilute the lensing signal (Appendix~\ref{sec:appendixA}).

\begin{figure}[]
    \includegraphics[width=1\columnwidth]{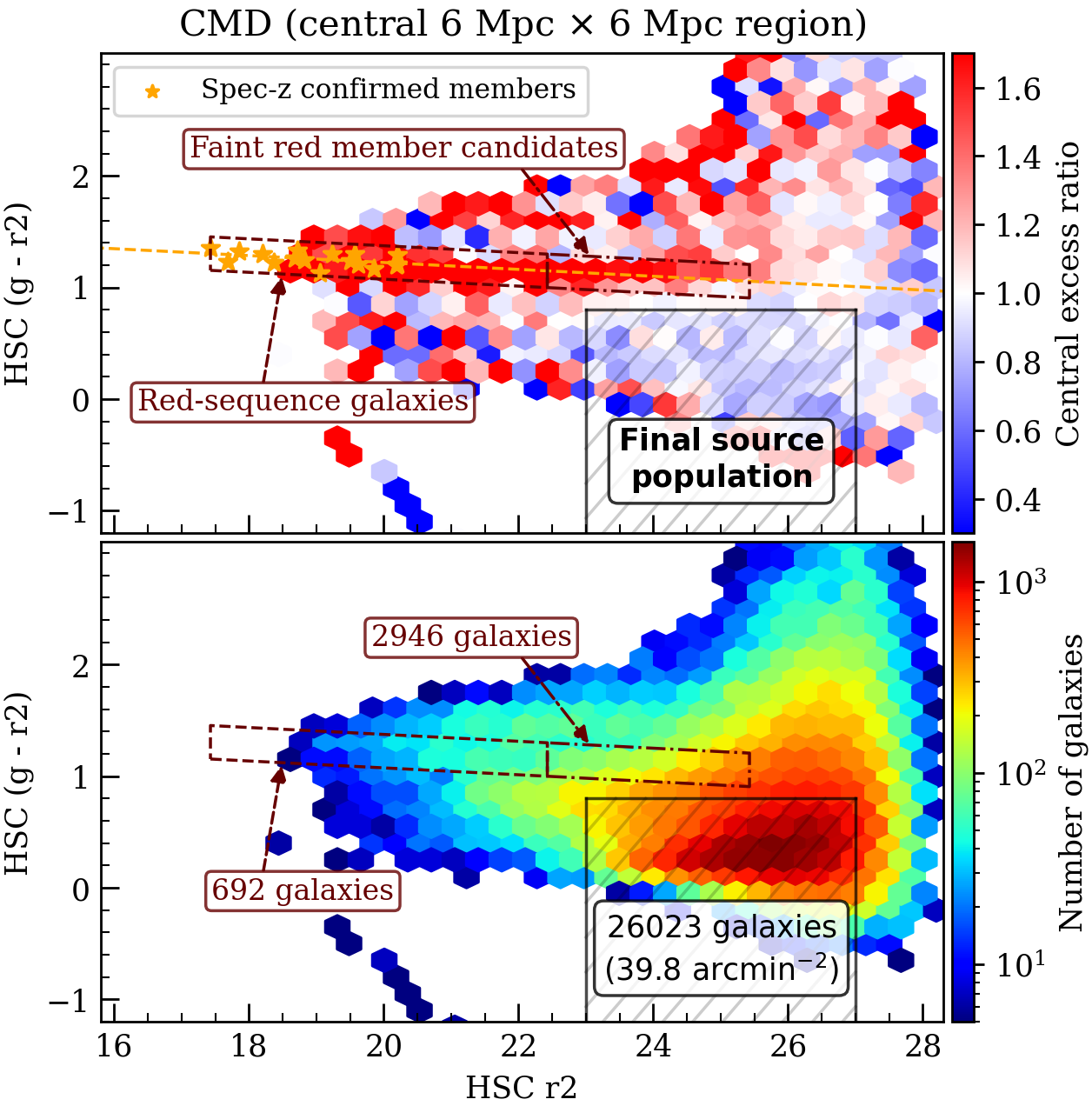}
    \caption{Color-magnitude diagram (CMD) of the PSZ2G181 field. Top: The color scale shows the central excess ratio of the galaxies, with the white color representing the uniform distribution and the blue and red represent the central deficiency and excess of the galaxies, respectively. For background sources, a central excess ratio less than unity is expected because of the obscuration by the cluster member galaxies. Bottom: Number of objects within each hexagonal bin as a color scale. Every bin contains at least five objects.}
    \label{fig:CMD_central_excess}
\end{figure}
Figure~\ref{fig:CMD_central_excess} shows the color-magnitude diagram (CMD) of the PSZ2G181 field.
The color scale of the upper plot is the central excess ratio of the number density within $R_{200c}$, defined as $n_{s,R_{200}} / n_{s,tot}$ where $n_{s,R_{200}}$ is the source density within $R_{\mathrm{200c}}$ and $n_{s,tot}$ is the global source density.
The central excess of the galaxies within $R_{200c}$ is clear along the red sequence down to the $r2$-band magnitudes of \mytilde25.
Therefore, we identify the red-sequence galaxies and the faint red member candidates along the cluster red sequence (Figure \ref{fig:CMD_central_excess}).
Following the conventional selection criteria for the red-sequence galaxies, we include those with $r2$-band magnitudes up to five magnitudes fainter than the BCG.
Faint red member candidates are galaxies with $r2$-band magnitudes between five and eight magnitudes fainter than the BCG.
Additionally, the galaxies that are redder than the faint red member candidates ($23 < r2 < 27$ and $g-r2 \ge 1.2$) also show a central excess, resulting in the dilution of the tangential shear signal (Figure~\ref{fig:tangential shear}).

Previous studies have reported that the faint blue population could be the major source of dilution \citep[e.g.,][]{Broadhurst.2005, Ziparo.2016, Medezinski.2018}.
However, in this case, we cannot find any hints of significant contamination from the faint blue galaxies.
Therefore, we identified the background source galaxies as those with $r2$-band magnitudes in the range of $23 < r2 < 27$ and $(g - r2)$ colors bluer than 0.8 (the hatched area in Figure \ref{fig:CMD_central_excess}), carefully accounting for both cluster member and foreground contaminations.

\subsubsection{Shape Selection Criteria}
The majority of the lensing signal comes from faint and small galaxies (which are typically found at higher redshifts and are less likely to be foreground sources).
However, the shape measure process becomes highly unstable for such faint and small galaxies.
Excessively including these galaxies may reduce the quality of the analysis, as their shape measurements are often unreliable.
To balance this, we selectively include faint galaxies while carefully filtering out those that fail to meet strict quality criteria.

First of all, the measurement errors for both ellipticity components ($e_1$ and $e_2$) should be smaller than their intrinsic dispersion \citep[shape noise; $\mathrm{\sigma_{SN}}=0.25$, e.g.,][]{Jee.2014_520, Hoekstra.2015}.
Also, the semi-minor axis $b$ should be larger than 0.4 pixels; shapes with smaller values exhibit the distinct ``bug pattern'' noted by \cite{Jee.2013} because of pixelation issues.
We additionally discarded objects with $e>0.9$ as they are usually the result of unstable fitting that approaches the theoretical limit ($e<1$).
Lastly, the $\chi^2$ fitting should be reliable ({\tt MPFIT STATUS} = 1) and {\tt SExtractor flag} should be less than 4.
After discarding these unreliable measurements, we obtained a mean source density of \mytilde39.8 arcmin$^{-2}$ in the entire field (\mytilde36 arcmin$^{-2}$ within the cluster's $R_{200c}$).
\begin{figure*}
    \includegraphics[width=0.95\textwidth]{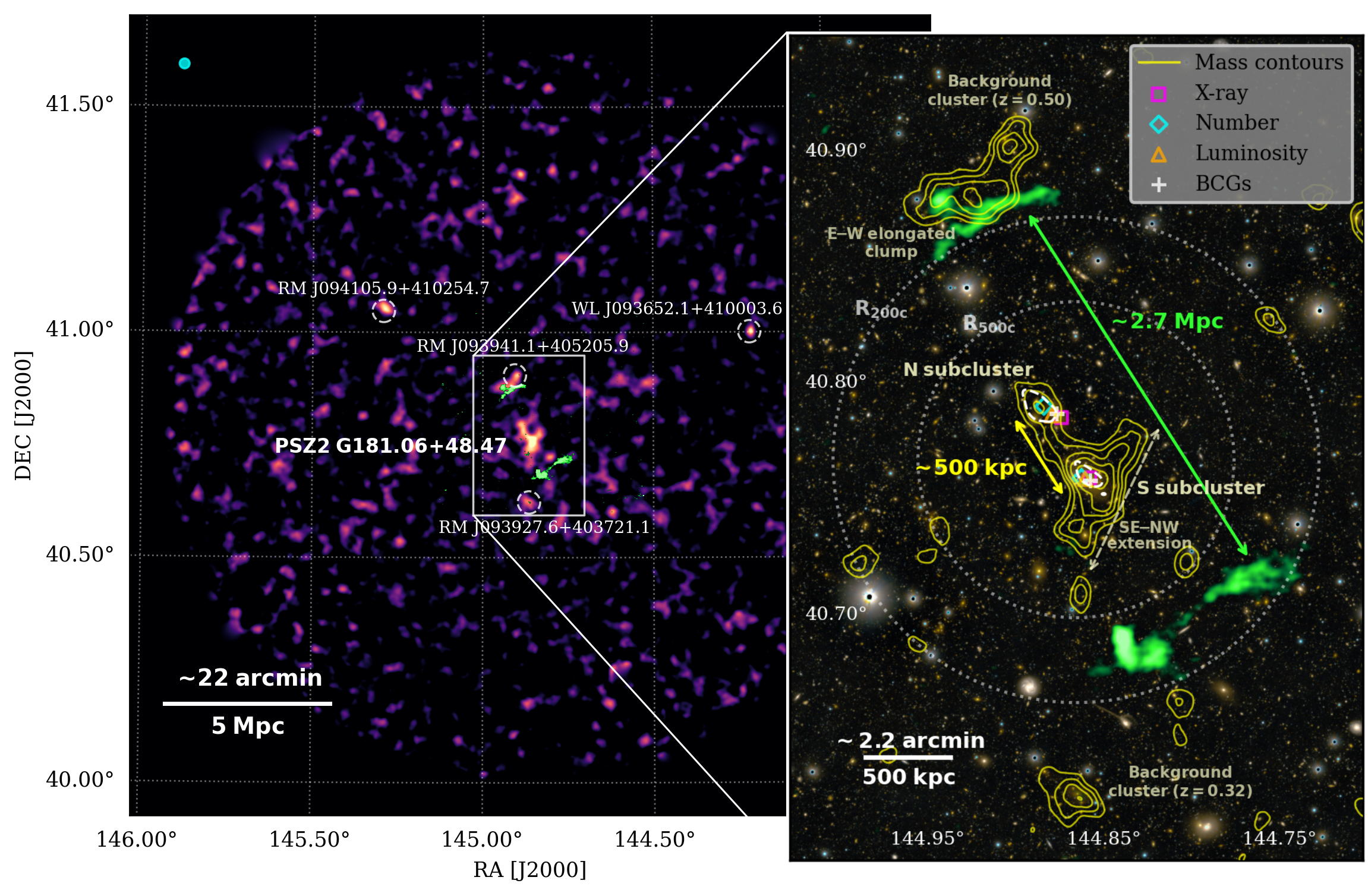}
    \caption{Reconstruction of the projected mass distribution of the PSZ2G181 field. The intensity of the LOFAR radio emission is represented in green. Left: Mass map of the full HSC coverage (\mytilde2 deg$^2$) using the \cite{Kaiser.1993} inversion method. Source galaxy ellipticities are averaged within an $r=40''$ top-hat kernel (cyan circle in the upper left). PSZ2G181 is detected in the central region, which is further resolved into the north and south subclusters. Other background cluster candidates within the field are marked as dashed circles (Appendix~\ref{sec:appendixB}). Right: Significance contours for the central 5~Mpc $\times$ 3~Mpc region from the 1000 bootstrap realizations using the {\tt FIATMAP} code \citep{Fischer.1997, Wittman.2006}. The spacing of the contour level is 0.5$\sigma$, with the lowest level of 2$\sigma$. The northern and southern subclusters are detected at \mytilde3.0$\sigma$ and \mytilde4.4$\sigma$ levels, respectively. The white dashed contours represent the 68\% confidence intervals of the mass centroids. The peaks in X-ray, red sequence number density, galaxy luminosity, and BCG for both subclusters are located within or near the 1$\sigma$ boundary of the mass centroids. $R_{200c}$ and $R_{500c}$ are centered on the barycenter of the two subclusters.}
    \label{fig:mass_full}
\end{figure*}

\subsection{Source Redshift Estimation}\label{sub:redshift estimation}
Since the two broadband filters are insufficient to constrain the photometric redshifts for individual sources, we calculated the effective source redshift using a control field.
We applied the same color-magnitude criteria to the GOODS-N galaxies after transforming the ACS magnitudes into the HSC magnitudes.
To account for differences in depth, we weighted the GOODS-N galaxies with the number density ratio in each magnitude bin between the GOODS-N and PSZ2G181 fields \citep[e.g.,][]{Finner.2017, Mincheol.2019}.
Since the WL signal is proportional to the angular diameter distance ratio $D_{ls}/D_s$ (Equation~\ref{eq:kappa}), we define its effective mean as 
\begin{equation}
    \left<\beta \right> = \left<max\left(0, \frac{D_{ls}}{D_s} \right) \right>,
\end{equation}
where we assigned zero weight to galaxies at redshifts lower than the cluster redshift.
We obtained $\left<\beta \right>=0.66$, corresponding to an effective source redshift of $z_\text{eff}=0.82$ with the width of $\left<\beta^2 \right>=0.49$.
Because of the non-linearity of the lensing kernel, the width of the redshift distribution should be accounted for when assuming the single source plane approximation.
We applied a first-order correction to the reduced shear \citep{Seitz.1997, Hoekstra.2000} as follows:
\begin{equation}\label{eq:first-order-correction}
    \frac{g'}{g} = 1 + \left(\frac{\left\langle \beta^2 \right\rangle}{\left\langle \beta \right\rangle^2}-1 \right) \kappa,
\end{equation}
where $g'$ and $g$ are the observed and true reduced shear, respectively.
From this equation, the observed shear $g'$ is scaled by a factor of $(1 + 0.12\kappa)$.

\section{Results}\label{sec:results}
\subsection{Two-dimensional Mass Reconstruction}\label{sub:mass map}
While the reduced shear is an observable quantity, we need to know shear $\gamma$ to obtain the surface mass density ($\kappa$) map.
The shear $\gamma$ is related to $g$ as $\gamma=g(1-\kappa)$, which can be converted into $\kappa$ via the following convolution \citep[][]{Kaiser.1993}:
\begin{equation}
    \kappa(\bs{\theta}) = \frac{1}{\pi} \int_{\mathbb{R}^2} D^*(\bs{\theta} - \bs{\theta '}) \gamma(\bs{\theta '}) d^2 \bs{\theta '},
\end{equation}
where $D^*(\bs{\theta}) = -1 / (\theta_1 - \bs{i}\theta_2)^2$ is the convolution kernel.
In the WL regime ($\kappa \ll 1$) where $\gamma$ approaches $g$, we can reconstruct the surface mass density using the reduced shear.

We generated the uncertainty (rms) map of the reconstructed mass map by bootstrapping the source catalog.
Next, we divided the median of the 1000 bootstrapped maps by the rms map to measure the statistical significance of the mass map.

\begin{figure*}
    \centering
    \includegraphics[width=0.9\textwidth]{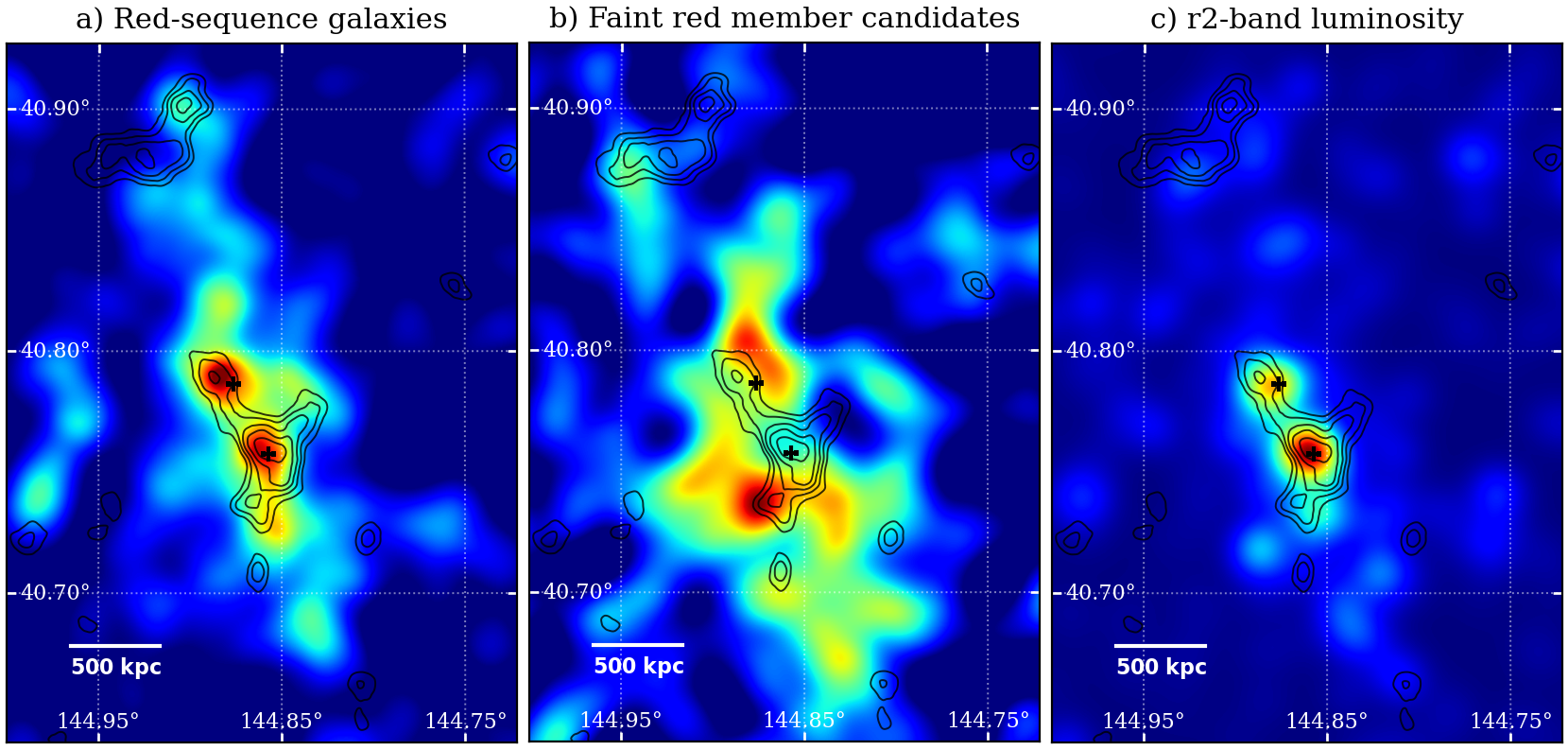}
    \caption{Comparison of the mass distribution of PSZ2G181 with the adaptively smoothed a) red-sequence galaxies and b) faint red member candidates number density maps, and c) the $r2$-band luminosity map smoothed with Gaussian kernel. The BCGs are marked as black crosses. All three maps show clear bimodalities. The two red sequence number density and luminosity peaks are cospatial with the mass peaks.}
    \label{fig:mass overlays}
\end{figure*}
Figure~\ref{fig:mass_full} shows the two-dimensional mass reconstruction of the PSZ2G181 field.
The mass reconstruction of the full HSC coverage (\mytilde2 deg$^2$) reveals several overdense regions, including four background cluster candidates (Appendix~\ref{sec:appendixB}).
An east-west elongated mass clump is detected near the NE relic, below the background cluster RM~J093941.1+405205.9.
We discuss more about this elongated mass in Section {\ref{sub:mass blob}}.
Among the overdense regions, PSZ2G181 is the most significant mass clump.
As shown in the zoomed-in image of the central 5~Mpc $\times$ 3~Mpc region, two separate mass peaks are detected for the northern and southern subclusters.
The direction of the separation coincides with the hypothesized merger axis defined in Figure~\ref{fig:color_xray_radio}.
Their physical separation is \mytilde500~kpc, much smaller than the relic separation (\mytilde2.7~Mpc).
The northern and southern subclusters are detected at a peak significance of \mytilde3.0$\sigma$ and \mytilde4.4$\sigma$, respectively.
While the northern mass peak exhibits a \mytilde100~kpc offset to the northeast from the northern BCG, the centroids of the BCG, galaxy luminosity, red sequence number density, and smoothed X-ray emission are either within or closely aligned with the 68\% confidence limit of the mass centroid.
X-ray peaks are determined from a Gaussian-smoothed X-ray map with a kernel of $\sigma = 10''$.

Figure~\ref{fig:mass overlays} presents the overlays of the mass contours with the following maps: a) red sequence number density, b) faint red member candidates number density\footnote{Identified in Figure \ref{fig:CMD_central_excess}.}, and c) $r2$-band luminosity map.
The two number density maps are adaptively smoothed with the minimal scale of $20''$, and the luminosity map is smoothed with the Gaussian kernel of $\sigma=30''$.
The red sequence number density and luminosity peaks are cospatial with their mass peaks.
The southern subcluster shows a mass extension that stretches along the SE-NW axis (right panel of Figure~\ref{fig:mass_full}) and is detected at the $\mytilde3 \sigma$ level.
Interestingly, the southern density peak of the faint red member candidates is located near the SE mass extension (Figure~\ref{fig:mass overlays}b), although this feature remains tentative.
We could not find any connection with the galaxies for the NW extension, although it could be related to a low-entropy gas plume located around this mass extension \citep[][]{Stroe.2025}.

The two number density maps exhibit a north-to-south elongation, parallel to the hypothesized merger axis (Figure \ref{fig:color_xray_radio}).
The distribution extends farther outward from the BCGs, with faint red member candidates reaching even greater distances.
This could be interpreted as `backsplash' galaxies\footnote{Backsplash galaxies are typically defined as galaxies located beyond the virial radius.} \citep[e.g.,][]{Gill.2005, Borrow.2023}.
The fraction of such galaxies is expected to increase following a cluster merger, becoming more widely dispersed over time \citep[e.g.,][]{Haggar.2020}.

\subsection{Mass Estimation}\label{sub:mass estimate}
\begin{figure}
    \centering
    \includegraphics[width=0.95\columnwidth]{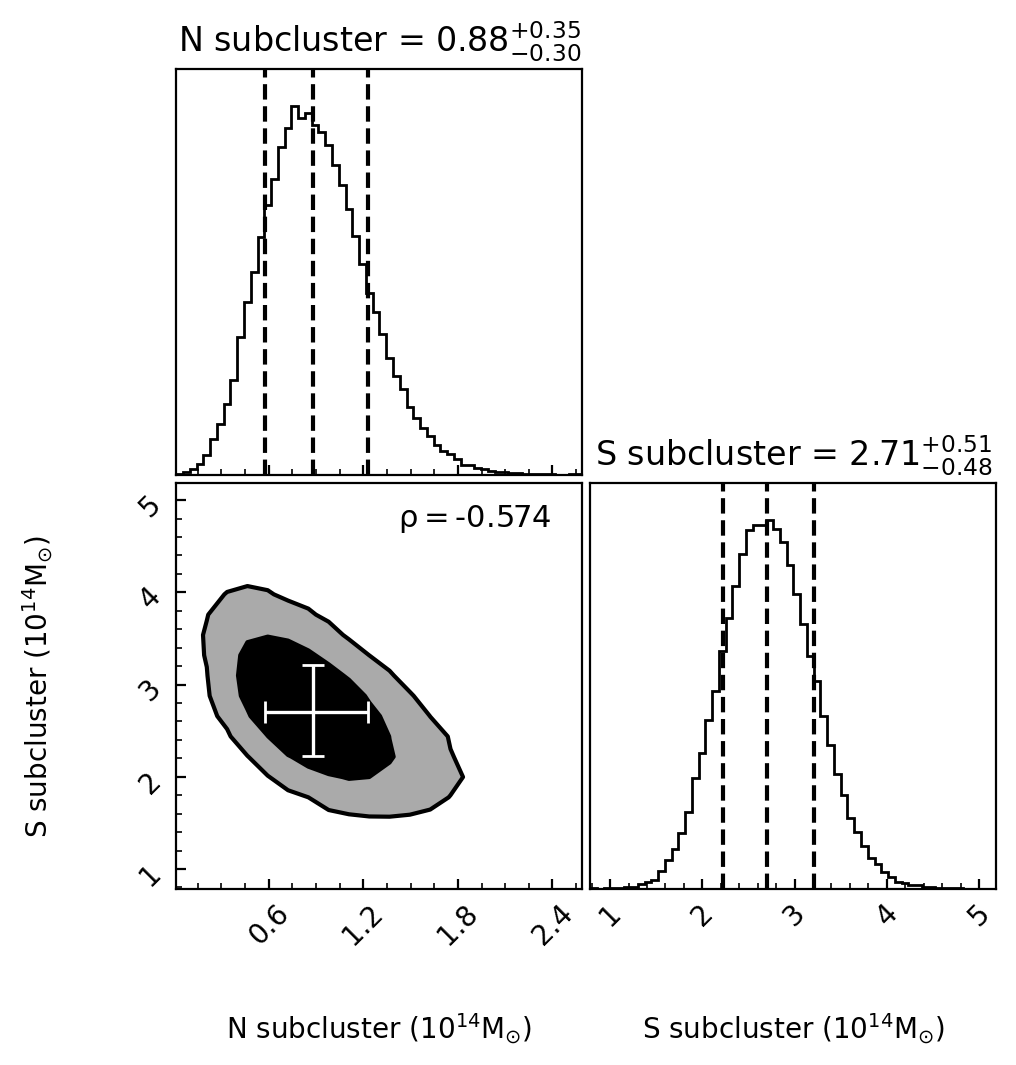}
    \caption{Mass estimation assuming the $M$--$c$ relation from \cite{Ishiyama.2021}. The black and gray contours represent the 68\% and 95\% confidence limits, respectively. The Pearson correlation coefficient $\rho$ is annotated. Since the two halos are closely separated, their masses are correlated. We note that the $M$--$c$ relation does not play a significant role in the mass estimation of this cluster.}
    \label{fig:mass_estimate1}
\end{figure}
For the mass estimation of PSZ2G181, we assumed the Navarro-Frenk-White \citep[NFW;][]{Navarro.1997} profiles for the northern and southern subclusters.
Since the separation between the two subclusters is relatively small (\mytilde500~kpc or \mytilde2$'$), we estimated the masses of the subclusters by fitting a superposition of the two NFW halos.
We assumed the mass--concentration ($M$--$c$) relation of \cite{Ishiyama.2021} to reduce the degeneracy between the two parameters in the NFW model.
We performed the Markov Chain Monte Carlo (MCMC) analysis with the following log-likelihood function \citep[][]{Mincheol.2019}:
\begin{equation}
    \mathcal{L} = - \sum_{i} \sum_{j=1, 2} \frac{[g_j^m(M_{\text{N}}, M_{\text{S}}, x_i, y_i) - g_j^o(x_i, y_i)]^2}{\sigma_{\text{SN}}^2 + (\delta e_{i,j})^2},
\end{equation}
where $M_{\text{N}}$ and $M_{\text{S}}$ are the masses of the northern and southern subclusters, respectively, and $g_j^m$ ($g_j^o$) is the $j^\text{th}$ component of the predicted (observed) reduced shear at the $i^\text{th}$ source galaxy position ($x_i$, $y_i$).
The shape noise $\sigma_\text{SN}$ is chosen to be 0.25, and $\delta e_{i,j}$ is the measurement error for the $j^\text{th}$ ellipticity component for the $i^\text{th}$ galaxy.
The predicted shear $g_j^m$ for every source galaxy position was calculated using the mathematical formulation of the NFW shear \citep[][]{Wright.2000}.
We assumed a flat prior of $10^{12}M_{\odot}< M_{200c} < 10^{16}M_{\odot}$.

In this study, we chose the BCG as the center for each subcluster for the mass estimation.
Because of recent merger activity, the location of the BCG might exhibit a slight offset from the center of gravitational potential.
This mis-centering effect could result in an underestimation of the masses \citep[e.g.,][]{Martel.2014, Zhang.2019}.
However, despite the \mytilde1$\sigma$ offset between the BCG and the mass peak of the northern subcluster, the centroids of both the red-sequence galaxies and the X-ray emission are located close to the BCG for both subclusters.
We excluded the sources within $r<40''$ from the center of each subcluster to further reduce the aforementioned mis-centering effect.
The mass of the northern (southern) subcluster is \mytilde18\% increased (\mytilde3\% decreased) if we position each halo at the mass peak, while the values are consistent within the 1$\sigma$ error range.

\begin{figure}
    \centering
    \includegraphics[width=0.95\columnwidth]{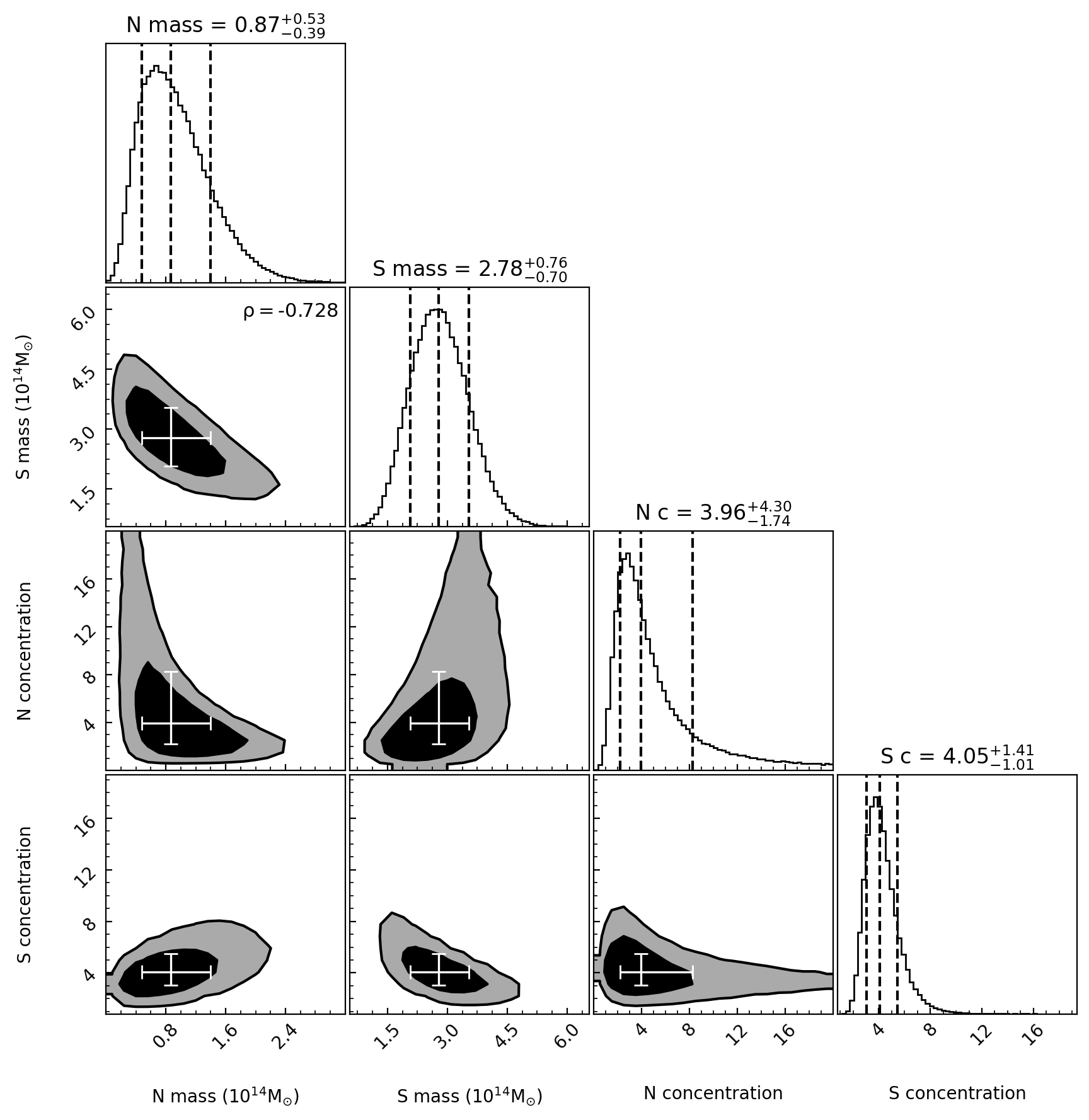}
    \caption{Mass estimation without assuming the $M$--$c$ relation. For the concentration, $0<c<20$ of flat prior is assumed. The masses of the northern and southern halos are consistent with the case for assuming the $M$--$c$ relation.}
    \label{fig:mass_estimate2}
\end{figure}
\begin{table*}
    \centering
    \caption{Mass Estimate of the PSZ2G181 Cluster}
    \label{tab:mass_estimate_is21}
    \begin{tabular*}{0.9\textwidth}{@{\extracolsep{\fill}} cccccc}
        \hline
        \hline
        \; &
        $M_{200c}$ & $M_{500c}$ & $R_{200c}$ & $R_{500c}$ &
        Mass peak \\
        \; &
        ($10^{14}M_{\odot}$) & ($10^{14}M_{\odot}$) &
        (Mpc) & (Mpc) & (RA, Dec) \\ 
        \addlinespace 
        \hline 
        Northern subcluster & 
        $0.88_{-0.30}^{+0.35}$ &
        $0.62_{-0.21}^{+0.25}$ &
        $0.85_{-0.11}^{+0.10}$ &
        $0.56_{-0.07}^{+0.07}$ &
        (144.8875$^{\circ}$, 40.7894$^{\circ}$) \\
        Southern subcluster & 
        $2.71_{-0.48}^{+0.51}$ &
        $1.87_{-0.33}^{+0.35}$ &
        $1.23_{-0.08}^{+0.07}$ &
        $0.80_{-0.05}^{+0.05}$ &
        (144.8589$^{\circ}$, 40.7591$^{\circ}$) \\
        Total system & 
        $4.22_{-1.00}^{+1.10}$ &
        $2.90_{-0.69}^{+0.75}$ &
        $1.43_{-0.12}^{+0.11}$ &
        $0.93_{-0.08}^{+0.07}$ &
        (144.8659$^{\circ}$, 40.7665$^{\circ}$)\footnote{Barycenter of the two mass peaks.} \\
        \hline
    \end{tabular*}
\end{table*}
\begin{table*}
    \centering
    \caption{Mass Estimate of the PSZ2G181 Cluster with Various $M$--$c$ Relations and without Assuming $M$--$c$ Relation}
    \label{tab:mass_estimate_mc}
    \begin{tabular*}{0.9\textwidth}{@{\extracolsep{\fill}} cccccc}
        \hline
        \hline
        $M$--$c$ relation &
        $M_{200c}^\text{N}$ & $M_{200c}^\text{S}$ & $M_{500c}^\text{N}$ & $M_{500c}^\text{S}$ & 
        $\rho$\footnote{Pearson correlation coefficient measured between the masses of the northern and southern subclusters.} \\
        \; &
        ($10^{14}M_{\odot}$) & ($10^{14}M_{\odot}$) &
        ($10^{14}M_{\odot}$) & ($10^{14}M_{\odot}$) & \\ 
        \addlinespace 
        \hline 
        \cite{Duffy.2008} & 
        $0.87_{-0.31}^{+0.36}$ &
        $2.80_{-0.51}^{+0.55}$ &
        $0.61_{-0.22}^{+0.25}$ &
        $1.94_{-0.35}^{+0.38}$ &
        -0.586 \\
        \cite{Dutton.2014} & 
        $0.88_{-0.29}^{+0.33}$ &
        $2.62_{-0.45}^{+0.48}$ &
        $0.62_{-0.20}^{+0.23}$ &
        $1.81_{-0.31}^{+0.33}$ &
        -0.556 \\
        \cite{Diemer.2019} & 
        $0.87_{-0.30}^{+0.35}$ &
        $2.72_{-0.48}^{+0.51}$ &
        $0.61_{-0.21}^{+0.25}$ &
        $1.88_{-0.33}^{+0.35}$ &
        -0.578 \\
        \cite{Ishiyama.2021} & 
        $0.88_{-0.30}^{+0.35}$ &
        $2.71_{-0.48}^{+0.51}$ &
        $0.62_{-0.21}^{+0.25}$ &
        $1.87_{-0.33}^{+0.35}$ &
        -0.574 \\
        Without $M$--$c$ relation & 
        $0.87_{-0.39}^{+0.53}$ &
        $2.78_{-0.70}^{+0.76}$ &
        $0.61_{-0.28}^{+0.37}$ &
        $1.92_{-0.48}^{+0.53}$ &
        -0.728 \\
        \hline
    \end{tabular*}
\end{table*}
Figure~\ref{fig:mass_estimate1} displays the posterior distributions of the masses of the northern and southern subclusters (see also Table~\ref{tab:mass_estimate_is21} for the best-fit values and the location of the mass centers).
We estimated the masses of the northern and southern subclusters to be $M_{200c}^\text{N}=0.88_{-0.30}^{+0.35} \times 10^{14} M_{\odot}$ and $M_{200c}^\text{S}=2.71_{-0.48}^{+0.51} \times 10^{14} M_{\odot}$, respectively.
Our results suggest that the PSZ2G181 cluster has undergone a major merger with a 3:1 mass ratio.
In many post-mergers, a dissociation between dark matter and ICM is frequently observed, and a less massive subcluster tends to have a brighter X-ray core \citep[e.g.,][]{Clowe.2006, Dawson.2012, Jee.2014, Cho.2022, Finner.2023}.
However, for this cluster, no significant dissociation is observed, and the more massive southern subcluster has a brighter X-ray core.
Since we suggest that this cluster is in a late merger phase (Section {\ref{sec:discussion}}), sufficient time might have passed for the ICM to stabilize within the gravitational potential center of each subcluster.

For the mass estimation of the total $M_{200c}$ ($M_{500c}$) of the system, we positioned the two NFW halos (N and S) in a three-dimensional grid and calculated the density for each pixel.
The density at each pixel is the sum of the contributions from both halos.
Starting from the barycenter of the two subclusters, we calculated the mean density within a spherical volume as a function of the increasing radius of the sphere.
Then, we defined the total $R_{200c}$ ($R_{500c}$) as the radius where the enclosed mean density is 200 (500) times the critical density at the cluster redshift.
The total mass of this system is $M_{200c} = 4.22_{-1.00}^{+1.10} \times 10^{14} M_{\odot}$ or $M_{500c} = 2.90_{-0.69}^{+0.75} \times 10^{14} M_{\odot}$.
This is marginally lower than the SZ mass \citep[$M_{\text{500,SZ}} = 4.2\pm0.5 \times 10^{14} M_{\odot} $;][]{Planck.2016}, while is in excellent agreement with the X-ray mass \citep[$M_{\text{500,X}} = 2.57_{-0.38}^{+0.37} \times 10^{14} M_{\odot} $;][]{Stroe.2025}.

$M$--$c$ relations are the average relations of a wide variety of halos, usually derived from the profile fittings using numerical simulations \citep[e.g.,][]{Duffy.2008, Ishiyama.2021}.
Therefore, assuming a specific $M$--$c$ relation could lead to a mass bias.
We repeated the mass estimation by applying four different $M$--$c$ relations, as well as without assuming any $M$--$c$ relation.
The results are summarized in Table~\ref{tab:mass_estimate_mc} and the posteriors without assuming the $M$--$c$ relation are displayed in Figure~\ref{fig:mass_estimate2}.
The masses for both the northern and southern subclusters are consistent regardless of the choice of the $M$--$c$ relation, even when the $M$--$c$ relation is not assumed.

\subsection{Remaining Systematics in Mass Estimation}
For the uncertainties of the mass estimation, only the shape noise and measurement errors are considered.
In this section, we discuss other systematics that could introduce additional bias to the mass estimation.

$\textit{NFW model bias}$.
Although the spherical halo models are widely used, the real 3D shapes of the DM halos are likely to be triaxial \citep[e.g.,][]{Corless.2009, Despali.2014, Lau.2021}.
A halo mass can be overestimated (underestimated) when the major axis of the triaxial halo is aligned with (perpendicular to) the line-of-sight direction.
Additionally, mass profiles of merging systems could deviate from the NFW profile because of complex dynamical states \citep[e.g.,][]{Wonki.2023}.
To estimate the amount of this model bias, we performed aperture mass densitometry \citep[AMD;][]{Fahlman.1994}.
AMD measures the projected mass enclosed within an aperture as a function of radius without assuming any parametric model.
We refer the readers to \cite{Clowe.2000} and \cite{Jee.2005} for details.
The marginal difference of the cumulative projected masses between the results from the NFW halo assumption (Table~\ref{tab:mass_estimate_is21}) and the AMD approaches is \mytilde13\% (\mytilde4\%) at $\mathrm{R_{200c}}$ ($\mathrm{R_{500c}}$), which is smaller than the statistical errors (\mytilde25\%) quoted in Table~\ref{tab:mass_estimate_is21}.

$\textit{Source redshift uncertainty}$.
The lensing signal depends on the relative geometry of the observer, lens, and background sources.
Therefore, imperfect source redshift estimation could lead to additional bias in the mass estimation.
When we apply the different control fields (four different catalogs used in Figure \ref{fig:lensing_SN}) for the source redshift estimation, the resulting scatter in the masses is \mytilde6\%.

$\textit{Large-scale structure}$.
Not only does the massive cluster change the shape of the background galaxies but the intervening large-scale structures along the line of sight also shear the galaxies.
Following the method in \cite{Hoekstra.2003}, the additional noise to the cluster-centric tangential shear due to the large-scale structures is $\sigma_{\text{LSS}}(1' < r < 30') = 0.0025\!-\!0.0045$, which is much smaller than the shape noise.

\subsection{Mass Clump Near the Northern Relic}\label{sub:mass blob}
Near the northern relic, an east-west elongated mass clump is detected in addition to the background cluster above it (RM~J093941.1+405205.9; Figure~\ref{fig:mass_full}).
The detection significance is at the $\mytilde3\sigma$ level. However, since we cannot find any bright galaxy counterparts, it is challenging to propose a plausible explanation for the nature of this clump.

One possibility is that this elongated mass clump is related to the background cluster RM~J093941.1+405205.9.
The distribution of red-sequence member candidates belonging to this background cluster is similar to the projected mass distribution in this region (second panel of Figure \ref{fig:background clusters}).
However, the X-ray emission from this cluster shows the unimodal distribution centered on the BCG candidate, with no elongation, weakening the connection between this cluster and the east-west elongated mass clump.

Another possibility would be that this mass clump is related to the PSZ2G181 cluster because the distribution of the faint red member candidates exhibits a weak concentration around this mass clump.
Further investigation of this region requires deep spectroscopy and high-resolution imaging data.

\section{Merger Scenario Reconstruction}\label{sec:discussion}
Our WL mass estimates suggest that the radio relics are located around $R_{200c}$ from the cluster center.
This makes PSZ2G181 a unique laboratory for studying cluster merger dynamics at the late phase of the merger and plasma acceleration in a low ICM density environment.
In this section, we define four configurations of the merger phase in time order and propose the most probable case based on our analyses.

\textit{Phase 1. Premerger.}
The two symmetric radio relics strongly indicate that the system is in a post-merger state, making this configuration highly unlikely.

\textit{Phase 2. Outgoing after the $1^\text{st}$ pericenter passage. \\}
If the system were in this merger phase, the shock speed should be unrealistically high to separate the two relics about \mytilde2.7~Mpc within a short timeframe (\mytilde0.2~Gyr, Figure \ref{fig:tsp hist}).

\textit{Phase 3. Returning after the $1^\text{st}$ apocenter passage.}
The most probable configuration.
We utilize idealized and cosmological simulations to validate this argument, as discussed in Section \ref{sub:returning}.

\textit{Phase 4. Outgoing after the $2^\text{nd}$ pericenter passage.}
Another possible but speculative configuration.
This merger phase requires multiple collisions that involve more than two halos.
We discuss this issue in Section \ref{sub:outgoing}.

\subsection{Returning Phase After the First Apocenter}\label{sub:returning}
We perform idealized simulations of cluster mergers using the adaptive mesh refinement code {\tt RAMSES} \citep[][]{Teyssier.2002}.
These simulations are not fine-tuned to exactly reproduce the observed multi-wavelength features.
Instead, we focus on the merger phase that can simultaneously reproduce the relic-to-relic (\mytilde2.7~Mpc) and the halo-to-halo (\mytilde500~kpc) separations.

\begin{figure}
    \centering
    \includegraphics[width=\columnwidth]{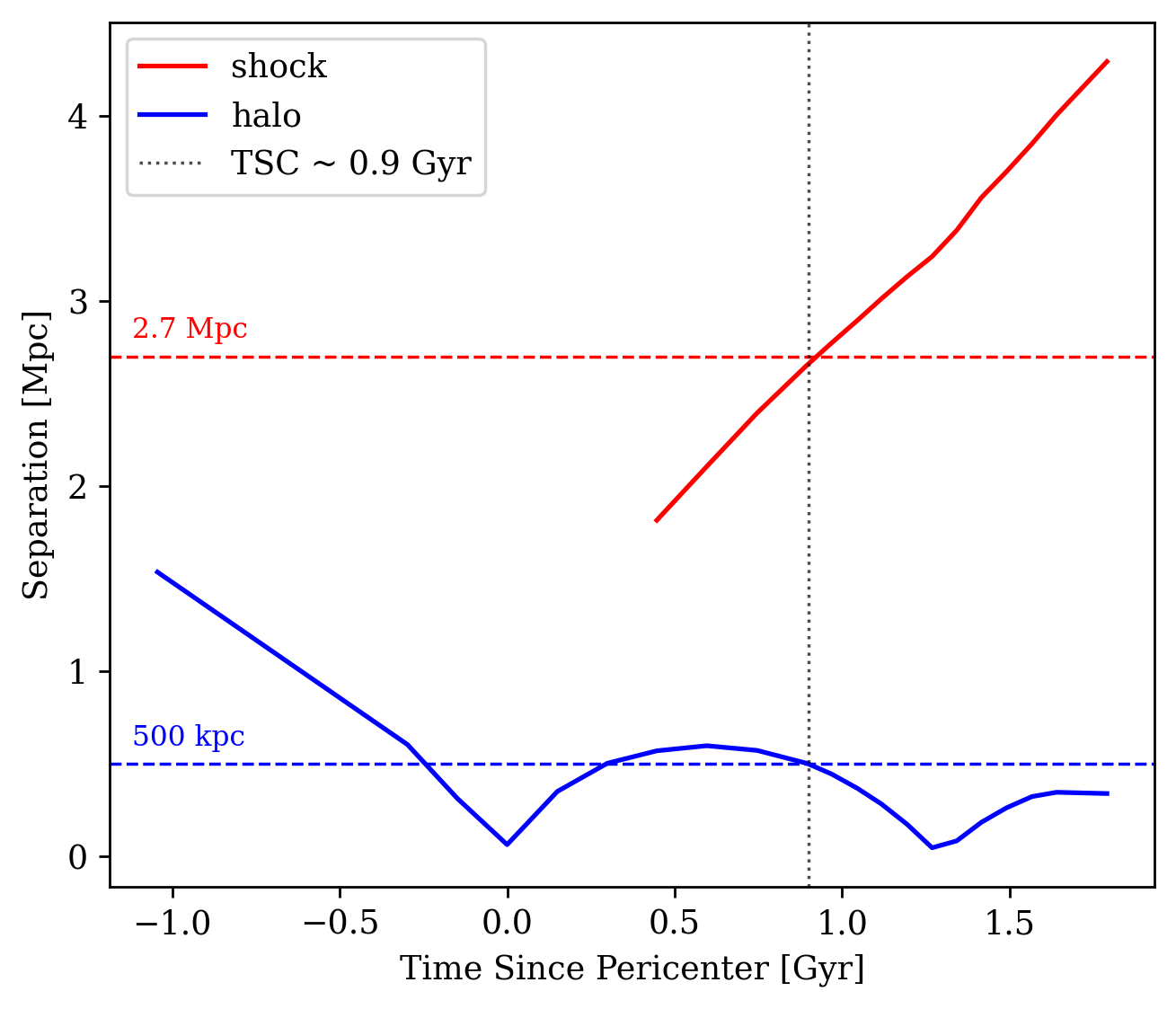}
    \caption{Time evolution of the halo separation (blue) and the shock separation (red). The dashed horizontal lines correspond to the observed halo-to-halo and shock-to-shock separations. Both halo and shock separations are reproduced \mytilde0.9~Gyr after the first pericenter passage in the idealized off-axis collision.}
    \label{fig:ideal sim}
\end{figure}
We use WL mass estimates to initialize the cluster masses.
We distribute the DM and ICM particles following the NFW profile and the modified beta profile \citep[][]{Vikhlinin.2006}, respectively, using the {\tt cluster\_generator}\footnote{https://github.com/jzuhone/cluster\_generator} package.
The clusters are placed with an initial separation of 2~Mpc and an initial relative velocity of 400$\mathrm{~km~s^{-1}}$.
We design the collision as an off-axis cluster merger to preserve the X-ray core after the collision as observed in PSZ2G181 (Figure \ref{fig:color_xray_radio}).
The simulation results in a pericenter distance of \mytilde0.1~Mpc.
The simulation adaptively refines the meshes to a minimum cell size of 4~kpc based on the local density and temperature gradient, allowing us to maintain a high resolution along the merger shocks in the low ICM density environment.
We follow \cite{Wonki.2020} to identify shockwaves and derive the relic-to-relic separation using the distance between the kinetic flux energy peaks of the merger shock pair.

Figure \ref{fig:ideal sim} shows the time evolution of the halo-to-halo and relic-to-relic separations in the simulated cluster merger.
After the first pericenter passage, the relic-to-relic separation increases linearly with time since the first pericenter ($\text{TSP}_{1^\text{st}}$), while the halo-to-halo separation oscillates until the two halos eventually merge into a single halo.
The observed separations are reproduced at $\text{TSP}_{1^\text{st}} \sim 0.9$~Gyr when the subclusters are returning for a second collision after having reached the first apocenter.
If we simulate the head-on cluster merger with zero impact parameter, the relic-to-relic separation is \mytilde2.0~Mpc when the halo-to-halo separation reaches \mytilde500~kpc after the first apocenter passage.

\begin{figure}
    \centering
    \includegraphics[width=\columnwidth]{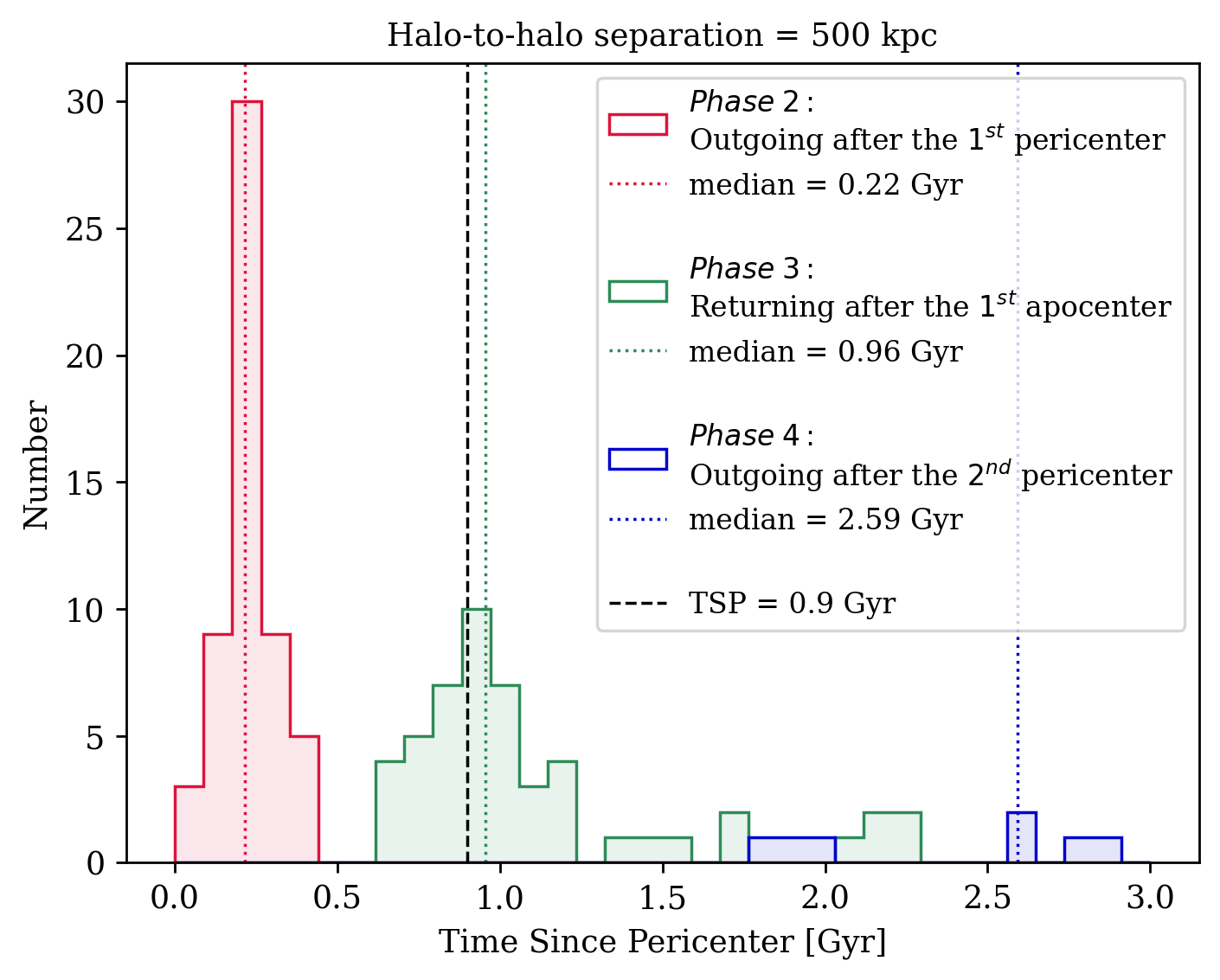}
    \caption{Distributions of the time since the first pericenter passage ($\text{TSP}_{1^\text{st}}$) of the 58 PSZ2G181 analogs in TNG-Cluster. In each case, $\text{TSP}_{1^\text{st}}$ is defined as the time when the halo-to-halo separation reaches 500~kpc. Many cluster mergers in TNG-Cluster show separations similar to that proposed in idealized simulations ($\text{TSP}_{1^\text{st}} \sim 0.9$~Gyr). Only seven systems have reached the halo separation larger than 500~kpc, which involve multiple collisions.}
    \label{fig:tsp hist}
\end{figure}
To account for realistic mass growth, we analyze the cluster mergers in the cosmological zoom-in simulation TNG-Cluster\footnote{https://www.tng-project.org/cluster/} \citep[][]{Nelson.2024}.
Among the \mytilde500 cluster merger events in TNG-Cluster \citep[][]{Wonki.2024}, we select 58 PSZ2G181 analogs with total mass ($3.2 \times 10^{14} M_{\odot} < M_{\text{200c,total}} < 5.2 \times 10^{14} M_{\odot}$) and the mass of the less massive subcluster ($0.6 \times 10^{14} M_{\odot} < M_{\text{200c,sub}} < 1.2 \times 10^{14} M_{\odot}$) are comparable to PSZ2G181.
Figure \ref{fig:tsp hist} shows the distribution of time since the first pericenter ($\text{TSP}_{1^\text{st}}$) of the PSZ2G181 analogs when the halo-to-halo separation is 500~kpc.
The median $\text{TSP}_{1^\text{st}}$ of the systems in \textit{Phase 3} is 0.96~Gyr, providing statistical support for our idealized simulations.

\subsection{Outgoing Phase After the Second Pericenter}\label{sub:outgoing}
In our idealized simulations, the subclusters cannot reach the observed separation (\mytilde500~kpc) after the second pericenter passage (Figure \ref{fig:ideal sim}).
Dynamical friction slows down the halos after the first pericenter passage, and the second collision starts from a free fall at the first apocenter separation.
As a result, the observed halo-to-halo separation cannot be reproduced at the outgoing phase after the second pericenter passage\footnote{Even with the doubled subcluster masses, the halo-to-halo separation cannot reach \mytilde500~kpc after the second collision.}.

\begin{figure}
    \centering
    \includegraphics[width=\columnwidth]{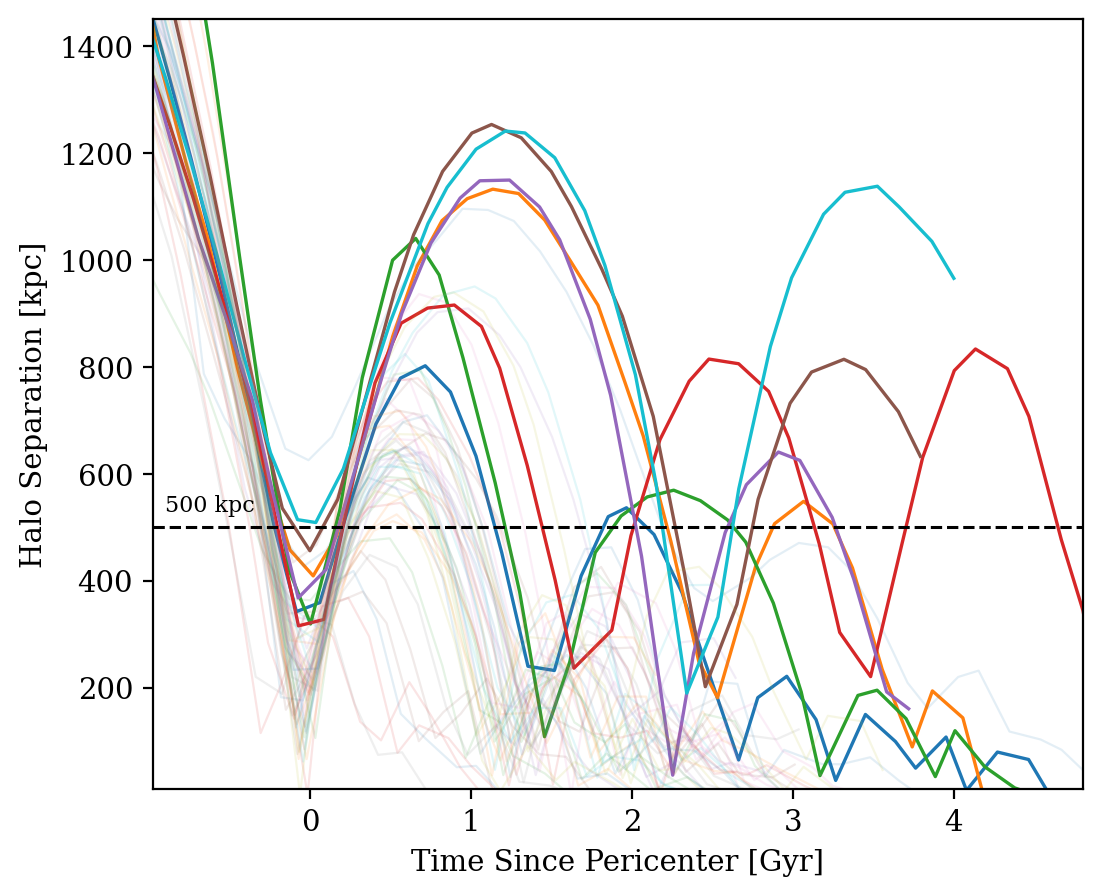}
    \caption{Time evolutions of the halo separations of the 58 PSZ2G181 analogs in TNG-Cluster. Seven systems that reached the halo separation greater than 500~kpc after the second pericenter passage are marked as thick lines. These systems involve the collision of more than two halos. The other 51 systems are marked with thin lines.}
    \label{fig:outgoing sep}
\end{figure}
However, unlike the idealized setup of the simulations, galaxy clusters accrete mass during the merger process, which can affect the halo separation at the late merger phase.
Figure \ref{fig:outgoing sep} illustrates the time evolution of the halo separations of the aforementioned PSZ2G181 analogs.
Among the 58 PSZ2G181 analogs in TNG-Cluster, seven systems have reached the halo separation greater than 500~kpc after the second pericenter passage (thick lines in Figure \ref{fig:outgoing sep}).
Interestingly, we find that all of these systems had experienced a collision with another cluster.
The additional collision provided gravitational excess, increasing the halo separation after the second core passage.
These seven systems have $\text{TSP}_{1^\text{st}} \gtrsim \text{2~Gyr}$ when the halo-to-halo separation reached 500~kpc after the second pericenter passage (blue in Figure \ref{fig:tsp hist}).

In the case of PSZ2G181, the mass distribution of the southern subcluster exhibits the SE-NW extension (Figure \ref{fig:mass_full}), and the SE extension shows a weak spatial correlation with the faint red member candidates (Figure \ref{fig:mass overlays}b).
If this substructure is a distinct halo that experienced a collision after the primary merger of PSZ2G181, this post-collision event could introduce additional gravitational forces.
These excess forces could alter the merger trajectories from those expected in a typical binary collision scenario (Figure \ref{fig:ideal sim}).
Interestingly, the southern relic has an inverted morphology \citep[][]{Rajpurohit.2025}, curving inward towards the cluster center.
The morphology of radio relics could be inverted if there are infalling tertiary subclusters \citep[][]{Wonki.2024}, which might support the multiple collision scenario.
X-ray discontinuities reported by \cite{Stroe.2025} may also result from the secondary collision involving the tertiary subcluster or the subclusters within PSZ2G181.

However, there are some unresolved issues that need to be addressed for \textit{Phase 4} to be the case.
First of all, $\text{TSP}_{1^\text{st}}$ is expected to be large ($\gtrsim$ 2~Gyr) to match the observed halo-to-halo separation, while the relic-to-relic separation would be too large at this moment.
Assuming a constant shock speed, the relics should be subsonic (\mytilde700~km~s$^{-1}$) to propagate \mytilde2.7~Mpc over 2~Gyr.
Furthermore, radio power is expected to rapidly drop after $\text{TSP}_{1^\text{st}} \sim 1$~Gyr \citep[][]{Nuza.2024}, potentially conflicting with the observed brightness of the PSZ2G181 relics.
Secondly, a collision involving a third halo would likely disturb the alignment of subclusters with the merger axis.
However, the BCGs, X-ray emission, galaxy density, and mass distribution align with the vector connecting the two relics.
Lastly, the X-ray emission shows a symmetric, bimodal distribution, with its peaks closely located near the two BCGs.
If the cluster were in \textit{Phase 4}, the X-ray emission would likely be dissociated from the mass peaks or BCGs, which contrasts with the observations in PSZ2G181.

Therefore, we infer that the binary collision cannot explain the observed halo separation after the second collision.
On the other hand, the multiple collision scenario is less probable due to the weak observational evidence.

\subsection{Large Viewing Angle Possibility}
\cite{Rajpurohit.2025} suggested that PSZ2G181 is projected with a large viewing angle ($\ge40^{\circ}$).
This implies that the deprojected halo-to-halo and relic-to-relic separations could be $\ge \text{700~kpc}$ and $\ge \text{3.5~Mpc}$, much larger than those observed.
According to our idealized simulations, the merger shocks reach the deprojected separation (\mytilde3.5~Mpc) at $\text{TSP}_{1^\text{st}} \sim 1.4$~Gyr, after having experienced the second collision.
However, as discussed earlier, it is more challenging to explain the deprojected halo separation (\mytilde700~kpc) after the second collision because of dynamical frictions.

One can suggest that the large viewing angle can be accommodated if we consider the misalignment between the planes of the halo and the radio relics.
An off-axis collision injects angular momentum into the cluster, triggering rotation in the halo distribution.
On the other hand, merger shocks are launched along the collision axis and are unaffected by rotation \citep[][]{Wonki.2020}.
Therefore, if the LOS vector aligns with the halo orbital plane and is perpendicular to the current halo separation vector, radio relics can be projected with a large viewing angle, while halos have minimal projection effects.
In such a configuration, $\text{TSP}_{1^\text{st}}$ is estimated as \mytilde1.6~Gyr near the second apocenter passage to explain the large deprojected relic-to-relic separation\footnote{Still, it is difficult to reproduce the \mytilde500~kpc of the halo separation after the second collision in a binary merger.}.

Another possible explanation is a high-velocity cluster merger.
With the initial velocity of \mytilde900$\mathrm{~km~s^{-1}}$, the halos take longer to return for a second core passage, and the shocks propagate at a higher speed.
Then the two halos and radio relics reach the deprojected separations at $\text{TSP}_{1^\text{st}} \sim$ 1.4~Gyr, during the returning phase after the first apocenter.
We note that the collision speed is \mytilde2500$\mathrm{~km~s^{-1}}$ for this scenario, which is unexpectedly high given the mass of PSZ2G181.

However, if the two relics are separated by \mytilde3.5~Mpc, they would be located far outside the cluster $R_{200c}$ (1.43~Mpc).
It would then be challenging to explain how the luminosity of the relics could be bright in such a low ICM density environment.
Additionally, radio relics appear broader and more diffuse when projected at a large viewing angle \citep[e.g.,][]{Skillman.2013}, which may contradict the observed thin relics.

\subsection{Summary of Merger Scenario Reconstruction}
In this section, we have explored the possibilities of four different merger phases in chronological order, as well as the possibility of a large viewing angle.
The pre-merger (\textit{Phase 1}) and outgoing phase after the first pericenter passage (\textit{Phase 2}) are easily rejected.
We found the returning phase after the first apocenter passage (\textit{Phase 3}) as the most probable case, supported by idealized and cosmological simulations.
The outgoing phase after the second pericenter passage (\textit{Phase 4}) may be possible if the system involves multiple collisions.
However, this case faces limitations, such as confirming tertiary substructures, requiring large $\text{TSP}_{1^\text{st}}$, accounting for the observed radio power and alignments among the cluster components, etc.

The large viewing angle of PSZ2G181 suggests a revised merger scenario with a longer $\text{TSP}_{1^\text{st}}$ (\mytilde1.5~Gyr).
However, the case of a large viewing angle presents unique challenges, such as verifying the misalignment between the planes of the halos and the radio relics or requiring an exceptionally high collision velocity.
Given the complexities in each scenario, further multi-wavelength observations and detailed simulations are required to better constrain the merger history of PSZ2G181.

\section{Summary and Conclusion}\label{sec:conclustion}
PSZ2~G181.06+49.47 is a merging galaxy cluster that exhibits bright and long double radio relics separated by \mytilde2.7~Mpc.
This separation is one of the largest among symmetric double radio-relic clusters, suggesting an unusually late phase of the cluster merger.
The symmetries of the two relics, two BCGs, ICM distribution, and galaxy density support the assumption of a binary collision model, enabling a reconstruction of the merger history.
We have presented the first WL analysis of PSZ2G181, characterized the dark matter structure, and suggested the merger scenario using numerical simulations.
Our new results are summarized as follows.

\begin{enumerate}

\item Our WL analysis using the HSC data detects two distinct subclusters in both the northern and southern regions, with a separation of \mytilde500~kpc.

\item The vector connecting the two subclusters is parallel to the vector connecting the two relics. The centroids of the X-ray, galaxy, and BCG are located within or closely aligned with the 68\% confidence limit of the mass centroid.

\item The masses for the northern and southern subclusters are $M_{200c}^{\text{N}}=0.9_{-0.3}^{+0.4} \times 10^{14} M_{\odot}$ and $M_{200c}^{\text{S}}=2.7_{-0.5}^{+0.5} \times 10^{14} M_{\odot}$, respectively.
The total $M_{200c}$ and $M_{500c}$ of the system is estimated to be $4.2_{-1.0}^{+1.1} \times 10^{14} M_{\odot}$ and $2.9_{-0.7}^{+0.8} \times 10^{14} M_{\odot}$, respectively.

\item We ran idealized simulations and found that an off-axis collision of 0.9 and $\mathrm{2.7 \times 10^{14} M_{\odot}}$ halos can reproduce the relic separation (\mytilde2.7~Mpc) and the halo separation (\mytilde 500~kpc) simultaneously.
They are returning toward each other \mytilde0.9~Gyr after the first pericenter passage, \mytilde0.3~Gyr after the first apocenter.
This is statistically supported by the cosmological zoom-in simulation TNG-Cluster.

\end{enumerate}

The low mass of PSZ2G181 and the large relic-to-relic separation suggest that radio relics could be bright and detectable even around the $R_{200c}$ of the cluster, where the ICM density is extremely low.
Therefore, the dissection of PSZ2G181 might give us a new insight into the particle acceleration in such a rare environment.
Our multi-wavelength analysis involving the X-ray \citep[][]{Stroe.2025} and radio \citep[][]{Rajpurohit.2025} studies serves as a key milestone toward a new understanding of the merging cluster physics.

\begin{acknowledgments}
We thank the TNG-Cluster team for allowing us to utilize the entire simulation dataset, which helps us to reconstruct the merging scenario.
M. J. Jee acknowledges support for the current research from the National Research Foundation (NRF) of Korea under the programs 2022R1A2C1003130 and RS-2023-00219959. 
A. Stroe gratefully acknowledges the support of a Clay Fellowship and the NASA 80NSSC21K0822 and $\mathit{Chandra}$ GO0-21122X grants.
W. Forman and C. Jones acknowledge support from the  Smithsonian Institution, the Chandra High Resolution Camera Project through NASA contract NAS8-03060 and NASA Grants 80NSSC19K0116, GO1-22132X, and ARO-21015X.
This work was supported by K-GMT Science Program (PID: GEMINI-KR-2022B-011) of Korea Astronomy and Space Science Institute (KASI). This research is based on data collected at the Subaru Telescope (PID: S22B-TE011-GQ), via the time exchange program between Subaru and the international Gemini Observatory, a program of NSF’s NOIRLab. The Subaru Telescope is operated by the National Astronomical Observatory of Japan. We are honored and grateful for the opportunity of observing the Universe from Maunakea, which has the cultural, historical, and natural significance in Hawaii. 
This paper makes use of LSST Science Pipelines software developed by the Vera C. Rubin Observatory. We thank the Rubin Observatory for making their code available as free software at \url{https://pipelines.lsst.io}.
\end{acknowledgments}

\facilities{
LOFAR, Subaru (Hyper Suprime-Cam), XMM
}


\appendix
\twocolumngrid

\section{Lensing Signal Dependence on Source Population}\label{sec:appendixA}

\renewcommand{\thefigure}{A\arabic{figure}}
\setcounter{figure}{0}
In addition to the qualitative discussion on the faint red member contamination in Section \ref{subsub:pho selection}, we investigated how the lensing signal depends on different color cuts of the background source population.
To quantify the lensing signal, we used the reduced tangential shear as:
\begin{figure}
    \includegraphics[width=\columnwidth]{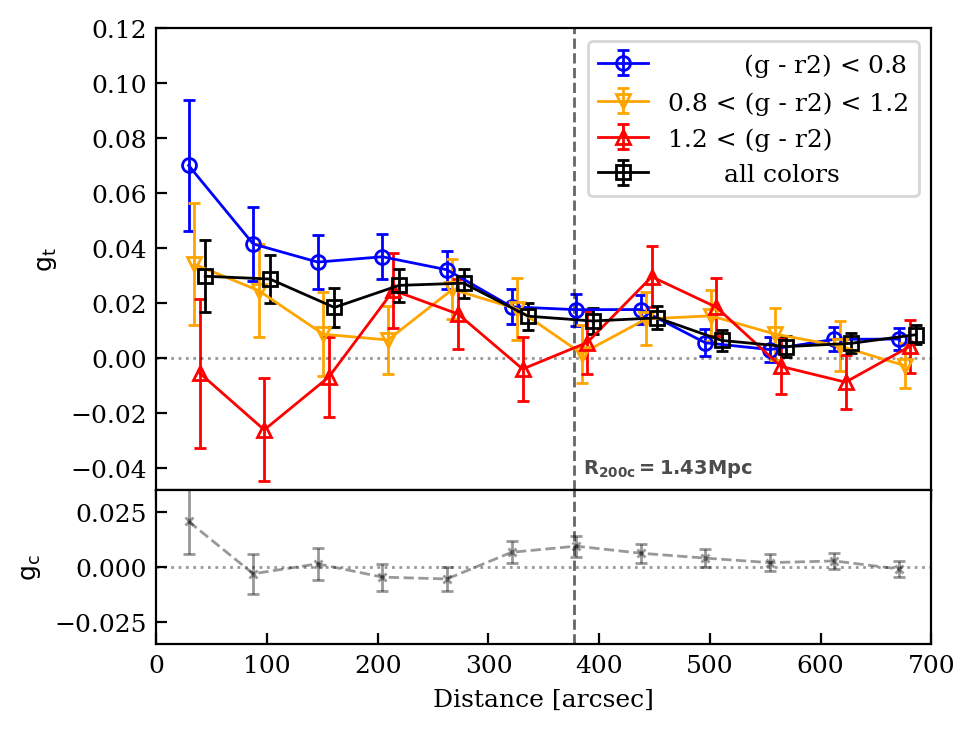}
    \caption{Top: cluster-centric tangential shear signal depending on the color selection window for the source population. Including the source galaxies redder than $(g-r)>0.8$ dilutes the lensing signal. Bottom: cross shear component which is consistent with zero.}
    \label{fig:tangential shear}
\end{figure}
\begin{equation}
    g_t = -g_1\;\text{cos}\;2\phi - g_2\;\text{sin}\;2\phi,
\end{equation}
where $g_1$ and $g_2$ are the two components of the reduced shear, and $\phi$ is the position angle with respect to the reference center.
Figure~\ref{fig:tangential shear} shows the tangential shear signal centered on the southern BCG for different color selection criteria.
We investigated four different populations: (1) our final selection ($g - r < 0.8$; blue), (2) those including faint red-sequence candidates ($0.8 < g - r < 1.2$; yellow), (3) red galaxies excluding the red sequence ($1.2 < g - r$; red), and (4) the full sample with no selection on their colors (black).
The magnitude range is fixed to $23 < r2 < 27$.
It is clear that including red source galaxies dilutes the lensing signal within the cluster $R_{200c}$.
This is possibly due to the contamination by faint member galaxies.
If we use all color ranges for the sources to maximize the $\mathrm{S/N_{lens}}$ (Figure~\ref{fig:lensing_SN}) without accounting for the faint cluster member contamination, the mass of the southern subcluster is \mytilde40\% underestimated.
We examined the cross shear (B-mode) which is a 45$^{\circ}$ rotation of the tangential shear and found it consistent with zero, indicating minimal residual systematics.

\section{Background Cluster Candidates within the PSZ2G181 field}\label{sec:appendixB}

\renewcommand{\thefigure}{B\arabic{figure}}
\setcounter{figure}{0}

\begin{figure*}
    \centering
    \includegraphics[width=0.95\textwidth]{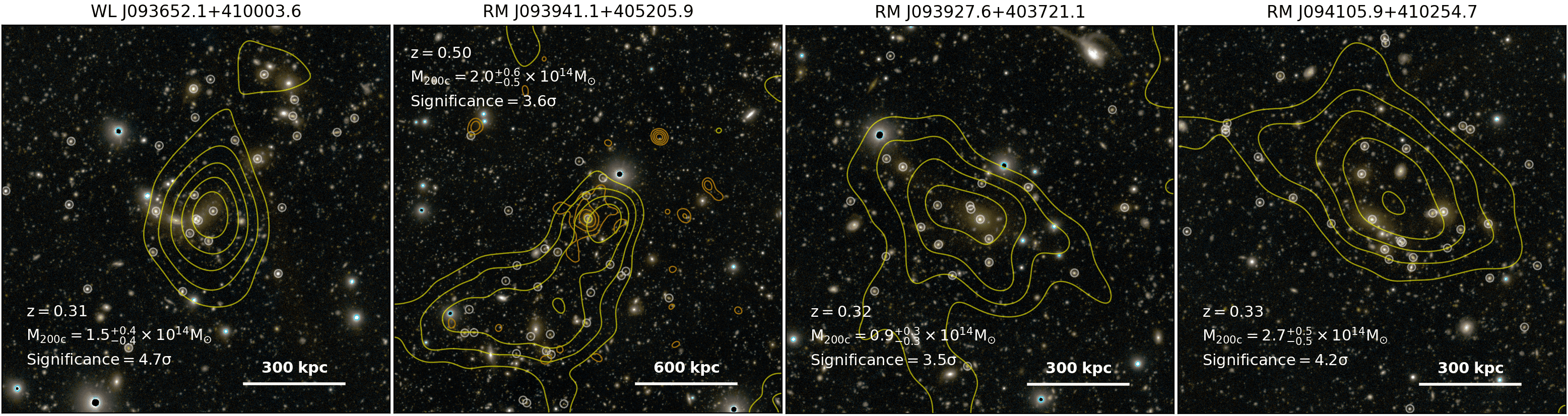}
    \caption{Three background cluster candidates within the \mytilde2~deg$^2$ HSC field of view centered at PSZ2G181.
    Yellow contours represent the convergence. White open circles represent the cluster red-sequence member candidates. For RM~J093941.1+405205.9 (second panel, one located above the northern relic), the $\mathit{Chandra}$ X-ray emission is marked as orange contours.}
    \label{fig:background clusters}
\end{figure*}
The large sky coverage and depth of our HSC observations allowed us to reconstruct the projected mass distribution across the $\mytilde\text{2 deg}^2$ PSZ2G181 field.
We identified four overdense regions in the projected mass map that have elliptical galaxy counterparts.
Three of them \citep[RM~J094105.9+410254.7, RM~J093941.1+405205.9, and RM~J093927.6+403721.1;][]{Rykoff.2014} were identified using the redMaPPer algorithm, and the other one (WL J093652.1+4 10003.6) was identified in \cite{Wen.2012}.
RM~J093941.1+405205.9 and RM~J093927.6+403721.1 were detected in \textit{XMM-Newton} observations \citep[C1 and C2 in][respectively]{Stroe.2025}.
Figure~\ref{fig:background clusters} shows the color-composite images of these background cluster candidates overlaid with mass contours.
Annotated masses are determined from the MCMC NFW halo fitting (Section \ref{sub:mass estimate}) assuming the $M$--$c$ relation from \cite{Ishiyama.2021}.
The center of the NFW halo is fixed to the BCG candidate.
Since RM~J093941.1+405025.9 was also covered in the $\mathit{Chandra}$ observation of PSZ2G181, we included the $\mathit{Chandra}$ X-ray surface brightness map for this cluster.
Note that the mass estimation of RM~J093941.1+405025.9 would be biased because of the puzzling mass clump located below it (Section \ref{sub:mass blob}).

\section{Strong-lensing Arc Candidates}\label{sec:appendixC}

\renewcommand{\thefigure}{C\arabic{figure}}
\setcounter{figure}{0}

\begin{figure*}
    \centering
    \includegraphics[width=0.85\textwidth]{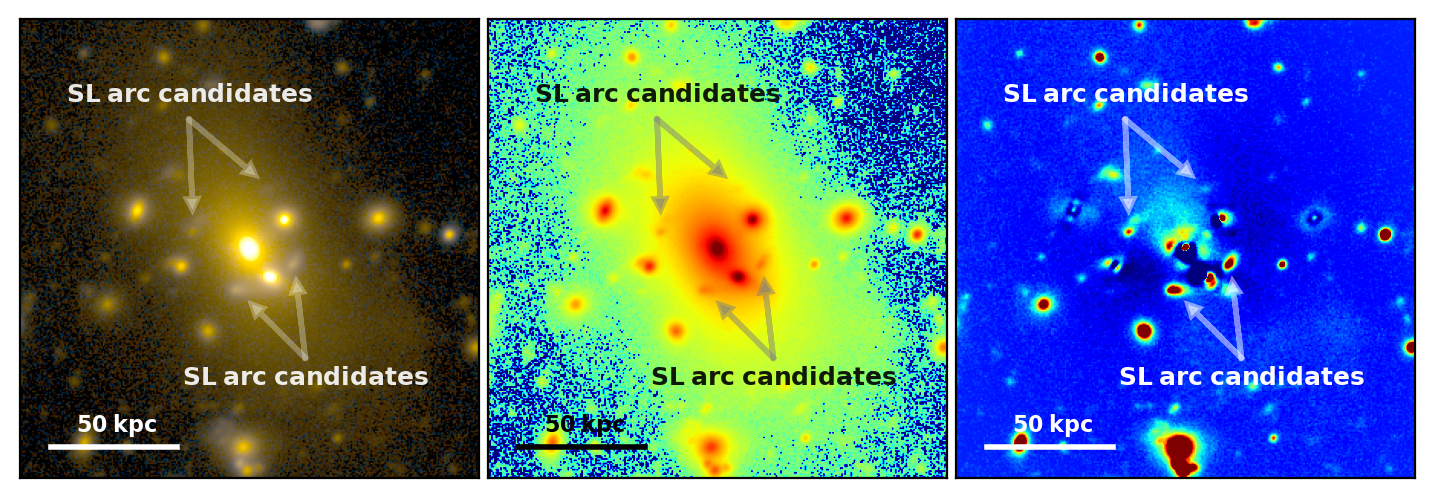}
    \caption{Strong-lensing arc candidates around the northern BCG. The left and middle panels show the color-composite and $r2$-band images around the northern BCG. The SL arc candidates are more clearly seen in the bright galaxy subtracted image (right) through Sersic profile fitting. The dynamic ranges of each panel are scaled differently for clarity. Arrows across the panels indicate the individual arcs at the same positions.}
    \label{fig:north SL}
\end{figure*}
Figure \ref{fig:north SL} displays the strong-lensing (SL) arc candidates around the northern BCG.
These candidates have different colors from the BCG and the red-sequence members, and they are curved outward of the BCG.

Since the resolution of HSC is insufficient to identify these candidates as multiple images, high-resolution space-based observations will be necessary to confirm them as SL images based on their morphology.
Additionally, spectroscopic observations are required to further validate these candidates as secure multiple images.
The mass of the northern subcluster is relatively low ($\mathrm{\mytilde1 \times 10^{14} M_{\odot}}$).
Therefore, if these images are indeed SL images, the northern halo should be more concentrated than expected from the $M$--$c$ relations.

\bibliography{reference}

\end{document}